\documentclass[linenumbers]{aastex631}

\usepackage{graphicx}	
\usepackage{amsmath}	
\usepackage{tablefootnote} 
\usepackage{placeins}

\begin{document}

\title{Exploring the evolution of a dwarf spheroidal galaxy with SPH simulations: II. AGN feedback}

\pagenumbering{gobble}

\author{Roberto Hazenfratz}
\affiliation{Núcleo de Astrofísica, Universidade Cidade de São Paulo \\
R. Galvão Bueno 868, Liberdade, 01506-000\\
São Paulo, Brazil}

\author{Paramita Barai}
\affiliation{Istituto Nazionale di Astrofisica (INAF) \\ Osservatorio Astronomico di Trieste (OATs) \\
Trieste, Italy}

\author{Gustavo A. Lanfranchi}
\affiliation{Núcleo de Astrofísica, Universidade Cidade de São Paulo \\
R. Galvão Bueno 868, Liberdade, 01506-000\\
São Paulo, Brazil}

\author{Anderson Caproni}
\affiliation{Núcleo de Astrofísica, Universidade Cidade de São Paulo \\
R. Galvão Bueno 868, Liberdade, 01506-000\\
São Paulo, Brazil}


\begin{abstract}

We investigate AGN feedback from an intermediate-mass black hole at the center of a dwarf spheroidal galaxy, by performing isolated galaxy simulations using a modified version of the GADGET-3 code. We consider Leo II (PGC 34176) in the Local Group as our simulation reference model. Beginning with black hole seeds ranging from $10^3$ to $10^6$ M$_{\odot}$, our simulations focus on comparing stellar-only feedback with AGN+stellar/SN feedback over 13.7 Gyr of galactic evolution. Our results indicate that a low-mass AGN in a dwarf galaxy influences the star formation history under specific physical conditions. While AGN feedback is generally negative on star formation, instances of positive feedback were also identified. Despite measurable effects on the evolution of the dwarf host galaxy, black hole seeds exhibited only marginal growth. We tested several physical scenarios as modified models in our simulations, primarily concerning the dynamics of the central black holes, which may wander within dwarf galaxies rather than being centrally located. However, none of these adjustments significantly impacted the growth of the black hole seeds. This suggests that intermediate-mass black holes may struggle to achieve higher masses in isolated environments, with mergers and interactions likely playing crucial roles in their growth. Nevertheless, AGN feedback exhibited non-negligible effects in our simulated dwarf spheroidal galaxies, despite the assumed dominant role of stellar feedback in the low-mass regime. 

\end{abstract}



\section{Introduction}

Hydrodynamical simulations of dwarf galaxies in the early Universe offer valuable insights into the fundamental physical processes underlying galactic formation. The Local Group of galaxies serves as an exceptional observational laboratory, providing valuable links to these simulations due to the great morphological variety and proximity of its satellite galaxies. Among the dwarf galaxies are the dwarf spheroidals (dSph), which represent the largest population within the Local Group. Investigating and observing these systems enables the analysis of advanced stages of primordial structures at low redshifts. An essential aspect in the study of the evolutionary journey of dSph galaxies is understanding the physical mechanisms responsible for depleting their gaseous content \citep{Mateo1998, Mcconnachie2012}. 

In massive galaxies, active galactic nuclei (AGN) play a significant role in shaping the formation and evolution of galaxies through their feedback mechanisms \citep{Silk1998}. This influence gives rise to observable trends such as the correlations between the central supermassive black hole (SMBH) and the bulge of the host galaxy \citep[e.g.,][]{Magorrian1998, Gebhardt2000}. However, at the low-mass end, reionization and stellar feedback are considered pivotal in suppressing star formation \citep[e.g.][]{Larson1974, White1991, Somerville2002}, with the role of AGN feedback initially presumed to be less significant in this mass regime. 

In recent years, there have been increasing observational evidence of AGN activity in different type of dwarf galaxies reported elsewhere, indicating a significant fraction (probably $> 50\%$) of galaxies with $10^9-10^{10}$ M$_{\odot}$ hosting black holes in the $10^4-10^5$ M$_{\odot}$ range \citep[e.g.][]{Lora2009, Jardel2012, Nucita2013, Manni2015, Reines2016, Baldassare2017, Nyland2017, Mezcua2020, Greene2020, Kimbrell2021, Yang2023}. Furthermore, recent works have suggested a significant influence from AGN feedback in dwarf galaxies, as opposed to the assumed unique influence from stellar feedback and environmental effects in the evolution of these systems \citep{Silk2017, Dashyan2018, Koudmani2022}. 

There are essentially three theoretical mechanisms proposed to explain the formation of seeds that could give rise to SMBHs observed ubiquitously in massive galaxies \citep[e.g.][]{Volonteri2010, Greene2012, Mezcua2017, Greene2020, Inayoshi2020, Connor2023}: 

(i) massive Population III star that ends its life and leaves a massive remnant of $\sim 10^2$ M$_{\odot}$.

(ii) formation through mergers within dense star clusters, resulting in remnants with masses ranging from $\sim 10^3-10^4$ M$_{\odot}$. 

(iii) direct collapse of atomic hydrogen, which may occur through a brief superstellar phase or directly into a black hole under specific conditions (such as exposure to Lyman–Werner photons, which can dissociate molecular hydrogen, thus preventing cloud fragmentation). This mechanism could produce remnants with masses ranging from $\sim 10^4-10^6$ M$_{\odot}$

One method to constrain the preferred mechanism for supermassive black hole formation, considering that all mechanisms could be at play, involves studying the occupation fraction of massive black holes (MBHs) in the local universe. For galaxies with masses $M_{\text{gal}} < 10^{10}$ M$_{\odot}$, the fraction appears to be lower than 1 \citep{Bellovary2019}. However, from an observational standpoint, investigating the active fraction of black holes as active galactic nuclei (AGNs) is a more feasible approach.

Another avenue to constrain the seeding mechanisms involves studying the smallest massive black holes and the less massive galaxies that host them, in order to approach their formation history and potential co-evolution. For this purpose, analyzing scaling relations such as $M_{BH}-\sigma_*$, $M_{BH}-M_{bulge}$ and $M_{BH}-M_*$, in this low-mass regime would provide further insights into SMBH seeding mechanisms \citep[e.g.,][]{Greene2020}. In this context, dwarf galaxies, the most prevalent type of galaxy in the universe, emerge as convenient laboratories for conducting such investigations.

Tracing the evolution of black holes in dwarf galaxies presents several challenges. Both observational and simulation-based studies have revealed that a significant fraction of these compact objects may be displaced from the center of their hosts. This displacement can be attributed to lower gravitational potentials, lower stellar densities, and the frequent disturbance caused by merger events within dwarf galaxies. The detection of these BH populations is further hindered by their low accretion luminosity and the lack of measurable dynamical effects on the stars within the host galaxy \citep{Greene2020, Mezcua2020, Reines2020, Sharma2020, Bellovary2021, Sharma2022}. Furthermore, distinguishing early seeding mechanisms using local observations can be challenging. This difficulty arises because diverse BH accretion histories can obscure early differences among them. To address this challenge, gravitational wave observations and studies of the luminosity function at high redshifts hold promise \citep{Bellovary2019, Greene2020}. 

The observation of AGNs in dwarf galaxies has spurred investigations into the impact of their feedback on galactic evolution through hydrodynamical simulations. Attempts to reproduce occupation fractions from these simulations have revealed a high level of discrepancy in the literature \citep[e.g.][and references therein]{Haidar2022}. The shallower gravitational potential wells within dwarf galaxies can amplify the impact of feedback mechanisms on the interstellar medium (ISM) and galactic evolution. This can manifest in the form of strong outflows capable of driving gas beyond the virial radius of the galaxy.

Until recently, AGN feedback has been overlooked in favor of the perceived dominance of stellar feedback in the low-mass regime of galaxies. However, recent evidence in the literature suggests that AGN feedback may indeed play a significant role in suppressing star formation in low-mass galaxies \citep[e.g.,][]{Silk2017, Bradford2018, Dashyan2018, Dickey2019, Manzano2019, Koudmani2019, Koudmani2022}. This emerging understanding underscores the importance of considering AGN feedback alongside other mechanisms in modeling the evolution of dwarf galaxies.

The galaxy chosen as the reference for this study is the dSph Leo II, first discovered in 1950 by \citet{Harrington1950}. With an estimated Galactocentric distance of $235.6^{+13.9}_{-9.14}$ kpc \citep{Li2021}, Leo II ranks among the most distant companions of the Milky Way, rendering it suitable for investigating internal feedback mechanisms with reduced influence from environmental effects, particularly in the more recent stages of galactic evolution.

While \citet{Spencer2017} argue that Leo II can reasonably be considered a satellite galaxy of the Milky Way based on its radial velocity and morphology, an alternative hypothesis posited by \citet{Coleman2007} suggests that Leo II may have evolved in isolation, given its low galactocentric component of radial velocity and the absence of evidence indicating disruption by tidal forces. Moreover, findings by \citet{Munoz2018} indicate that Leo II is the roundest dwarf galaxy observed in their MegaCam survey ($\epsilon = 0.07 \pm 0.02$), exhibiting a regular morphology devoid of discernible signs of tidal features.

Despite the emergence of several related works in recent literature, the exact role of AGNs in suppressing star formation within dwarf galaxies remains an open question. Within this context, the objective of this study is to investigate the influence of a dwarf AGN and its feedback on the evolution of a dwarf spheroidal galaxy resembling Leo II, particularly in relative isolation from the massive hosts of the Local Group.

In a broader sense, the isolated simulations conducted in this study may yield optimal parameters for cosmological and zoom-in simulations by exploring the parameter space of subgrid models of AGN feedback in the low-mass regime of classical dwarfs. While some prior works, such as in \citet{Dubois2015,Habouzit2017} and \citet{Koudmani2022}, have focused on understanding how supernova (SN) feedback influences black hole (BH) growth in dwarf galaxies, our study takes a different approach. We aim to investigate the influence of AGN feedback on the evolution of a reference dwarf galaxy using subgrid models for accretion and energy input into the interstellar medium. To accomplish this, we consider a fiducial model derived from previous research that focused on examining the influence of stellar feedback on the evolution of the dwarf spheroidal galaxy Leo II in \citet{Hazenfratz2024}, which was the first paper of this series. This second paper explores AGN feedback from a putative intermediate-mass black hole, analyzing its growth and impacts on the evolution of our benchmark galaxy.

\section{Numerical Methods}

The hydrodynamical simulations aimed at investigating AGN feedback in a dwarf spheroidal galaxy were conducted using a modified version of the TreePM (particle mesh) – smoothed particle hydrodynamics (SPH) code GADGET-3. These simulations are part of a broader project focused on exploring various processes associated with the formation and evolution of dwarf spheroidal galaxies within the Local Group. The general objective of the project is to analyze the role and interplay of internal feedback mechanisms within galaxies and environmental influences. 

In the simulations of this work, we adopt initial conditions consistent with those described in \citet{Hazenfratz2024}, where the distribution of dark matter and the initial distribution of gas follow a Hernquist profile. Moreover, we utilize the adaptive gravitational softening length method developed by \citet{Iannuzzi2011}.

Radiative cooling processes are implemented for helium and hydrogen and combined with lines for radiative cooling for metals C, Ca, O, N, Ne, Mg, S, Si and Fe \citep{Tornatore2007, Wiersma2009}. The cooling tables represent a gas exposed to a redshift-dependent UV/X-ray background radiation from quasars and galaxies, based on the model of \citet{Haardt2001}, alongside the redshift-dependent cosmic microwave background radiation. The gas is assumed to be optically thin and in ionization equilibrium. 

In the GADGET-3 code, a black hole is represented as a collisionless particle, characterized by two distinct masses. The subgrid mass (\textit{$M_{BH}$}) is employed to calculate pertinent physics, including the Bondi rate (eq.~\ref{eq:bondi}), and can be considered the true mass of the black hole. Meanwhile, the dynamical mass (\textit{$m_{BH,dyn}$}) governs non-AGN physics within the code, such as gravitational interactions, and is intricately linked to the resolution of the simulations. At $t = 0$, a black hole is initialized within the galaxy's central collisionless particle. The mass of this seed varies from $M_{BH} = 10^3 - 10^6$ M$_{\odot}$ across the simulations.

\subsection{Stellar feedback}

In this study, we implement star formation and feedback mechanisms tailored to resolved scales (approximately $\sim 10^4$ M$_{\odot}$ for gas). The star formation process follows the effective subgrid scheme proposed by \citet{Springel2003}, where the ISM is depicted as a fluid composed of cold clouds ($T_c\sim1000$ K) confined within and in pressure equilibrium with a hot ambient gas ($T_h\sim10^5-10^7$ K). The model accounts for three fundamental processes governing mass exchange between phases: star formation, evaporation of cold clouds due to supernova feedback, and growth of cold clouds through radiative cooling. This approach enables the self-regulation of star formation by incorporating stellar winds.

Kinetic stellar feedback, in the form of winds from supernovae, is implemented using the energy-driven prescription and chemical evolution model detailed in \citet{Tornatore2007}. This model explicitly incorporates stellar lifetimes to determine the release of metals and energy. Each star particle is regarded as a simple stellar population, with its mass evolving according to mass-dependent stellar lifetimes based on a chosen initial mass function (IMF), while also considering mass losses. In this study, we adopt the stellar IMF proposed by \citet{Chabrier2003} within the mass range [0.1, 100] M$_\odot$.

\subsection{AGN feedback}

We incorporated both thermal \citep{Springel2005, Booth2009} and kinetic \citep{Barai2014} AGN feedback to investigate the impact of a putative intermediate-mass black hole (IMBH) activity on the evolution of an isolated dwarf spheroidal galaxy. Given the current challenges in observing dwarf galaxies beyond the Local Group, robust constraints on dwarf AGN models remain elusive. Thus, we utilize constraints derived from previous studies employing the same AGN feedback models for larger galaxies \citep{Barai2014, Barai2019} as an upper limit for setting up the simulations (see Table~\ref{tab:simul_table}). Additionally, we consider references from prior research on dwarf spheroidals from \citet{Lanfranchi2021} to inform our approach.

The black hole feeding is based on the mass accretion model of \citet{hoyle1939}, \citet{bondi1944} and \citet{bondi1952}:

\begin{equation}
    \label{eq:bondi}
    \dot M_{Bondi} = \alpha \frac{4\pi G^2 M_{BH}^2 \rho_{\infty}}{(c_{s,\infty}^2 + v^2)^{3/2}}
\end{equation}

where, in the original formulation, $\rho_{\infty}$ represents the gas density far from the BH (at infinity), $c_{s,\infty}$ is the gas sound speed far from the BH, \textit{v} is the BH velocity relative to the distant gas. The parameter $\alpha$ is introduced as a numerical correction factor to account for the lack of resolution of the Bondi radius and ISM details in our N-body simulations.

The gas properties in Equation \ref{eq:bondi} ($\rho_{\infty}$, $c_{s,\infty}$) are estimated by smoothing at the black hole location according to the resolution scale. This procedure typically yields lower densities compared to cases where the Bondi radius scale is resolved. Moreover, the accreted gas is multiphase, but the cold phase is not resolved in detail in our simulations, leading to a reduced estimated accretion rate. As a result, corrections are often made by setting $\alpha \sim 100$ \cite[e.g.,][]{Springel2005} in many studies involving disk-like galaxies. However, it remains uncertain whether this factor is suitable for simulations of dwarf galaxies, necessitating further investigation (see Section \ref{accretion_fac}).

The BH accretion is limited by the Eddington rate

\begin{equation}
    \label{eq:max_accretion}
    \dot M_{BH} = min(\dot M_{Bondi}, \dot M_{Edd}) \
\end{equation}

where $\dot M_{\text{Edd}}$ represents the Eddington accretion rate, which can be expressed in terms of the Eddington luminosity

\begin{equation}
   \label{eq:edd_lumin}
   L_{Edd} = \frac{4\pi G M_{BH} m_p c}{\sigma_T} = \epsilon_r \dot M_{Edd} c^2
\end{equation}

A fraction of the rest-mass energy accreted by the BH is radiated away according to

\begin{equation}
   \label{eq:rad_lumin}
   L_r = \epsilon_r \dot M_{BH} c^2
\end{equation}

In all simulations, the adopted value $\epsilon_r \sim 0.1$ corresponds to a commonly used mean value for radiatively efficient accretion onto black hole for a variety of physical conditions \citep[e.g.,][]{Shakura1973, Narayan1995}. This value is based on the standard thin accretion disk model, which provides a reasonable approximation for many astrophysical scenarios.

A fraction of the radiated energy is coupled back to the surrounding gas according to

\begin{equation}
   \label{eq:feed_energy}
   \dot E_{feed} = \epsilon_f L_r = \epsilon_f \epsilon_r \dot M_{BH} c^2
\end{equation}

The subgrid model for the thermal AGN feedback is based on the prescriptions of \citet{Springel2005}, where $\dot E_{\textit{feed}}$ is distributed isotropically to heat up the gas around the BH. The temperature of the gas particles is increased by an amount scaled by the SPH kernel weights. 


We employ a kinetic AGN feedback model, which is characterized by two free parameters: the feedback efficiency $\epsilon_\textit{f}$ and the AGN wind velocity $v_w$. While the subgrid model proposed by \citet{Barai2014} adopts a fixed value for the wind velocity as a simplifying assumption, in practice, the AGN wind velocity should be self-regulated. The energy-conservation equations can be expressed using either the kinetic energy or momentum of the affected gas, giving rise to two distinct AGN wind formalisms: energy-driven or momentum-driven wind. Evidence from cosmological simulations suggests that the former formalism is likely more suitable for low-mass galaxies \citep[e.g.,][]{Murray2005, Murray2009}. In this approach, the kinetic energy carried away by the wind is equated to the AGN feedback energy:

\begin{equation}
   \label{eq:energy-driven}
   \frac{1}{2} \dot M_w v_w^2 = \dot E_{feed} = \epsilon_f \epsilon_r \dot M_{BH} c^2
\end{equation}

Rearranging equation~\ref{eq:energy-driven} to obtain the gas outflow rate gives

\begin{equation}
   \label{eq:outflow-rate}
   \dot M_w = 2 \epsilon_f \epsilon_r \dot M_{BH} \frac{c^2}{v_w^2}
\end{equation}

\subsection{Numerical implementation of AGN feedback}

We opted to explore the mass range of the intermediate mass black hole from $10^3$ to $10^6$ M$_{\odot}$, with the latter serving as an upper limit between the intermediate and supermassive black hole regimes. We excluded $10^2$ M$_{\odot}$ from our tests as it would result in a disparity of two orders of magnitude between the initial black hole dynamical and subgrid masses, considering our gas particle resolution is approximately $\sim 10^4$ M$_{\odot}$.

At each timestep $\Delta t$, the black hole undergoes growth according to the Bondi accretion rate in equation \ref{eq:bondi}. Here, the subgrid mass increases by $\dot M_{\textit{BH}}\Delta t$ without affecting the surrounding gas distribution. The dynamical mass $m_{\textit{BH,dyn}}$ remains constant until an accretion event, which typically happens when $M_{\textit{BH}} \geq m_{\textit{BH,dyn}}$. Upon accreting a gas particle, it is eliminated, and its mass is incorporated into $m_{\textit{BH,dyn}}$, thus conserving the dynamical mass within the computational volume.

Gas accretion follows a stochastic methodology, whereas the subgrid black hole mass $M_{\textit{BH}}$ increases smoothly. The kernel employed for computing quantities around the BH maintains the same shape as that used in SPH calculations, with four times the number of neighbors (4 x 32). The size of the kernel, determined by the BH smoothing length $s_{\textit{BH}}$, is computed through the implicit solution of the equation

\begin{equation}
   \label{eq:s_bh}
   \frac{4}{3} \pi s_{BH}^3 \rho _{BH} = M_{ngb}
\end{equation}

where $\rho _{\textit{BH}}$ is the kernel estimate of the gas density at the position of the BH and $M_{ngb}$ is the mass of $\sim 4 \times 32$ neighbouring gas particles. 

A probabilistic criterion is also employed to distribute AGN kinetic feedback energy among the surrounding gas particles. These particles are stochastically selected and propelled as AGN winds with a velocity boost $\textbf{v}_w$. The probability of a gas particle being boosted is calculated by

\begin{equation}
   \label{eq:prob_kick}
   p_i = \frac{w_i \dot M_w \Delta t}{\rho_{BH}}
\end{equation}

where $w_i = W(|r_{\textit{BH}} - r_i|, s_{\textit{BH}})$ is the SPH kernel weight of the gas particle with respect to the BH, and $\dot M_w$ is the gas outflow rate as expressed in eq.~\ref{eq:outflow-rate}. Random numbers $x_i = [0,1]$ are uniformly distributed for each gas particle. When $p_i > x_i$, a gas particle with velocity $\textbf{v}_{old}$ undergoes a velocity boost:

\begin{equation}
    \label{eq:vel_kick}
    \textbf v_{new} = \textbf v_{old} + v_w \textbf {\^x}
\end{equation}

In the simulations conducted in this study, the AGN wind particles are never hydrodynamically decoupled. Moreover, the fraction of the accreted rest-mass energy injected into the interstellar medium is assumed to be independent of both the environment and the accretion rate.


\subsection{Overview of the simulations}

Table~\ref{tab:config_table} provides an overview of the general configuration and initial conditions for the simulations conducted using the GADGET-3 code, spanning approximately 13.7 Gyr of galactic evolution. The virial velocity value was derived from the lower limit of the circular velocity estimated by \citet{Strigari2007}. It is noteworthy that the utilized value differs from the approximate 17 km s$^{-1}$ reported by the authors. The primary aim was to reproduce the mass of the galaxy estimated at $z=0$, which closely aligns with the dark matter halo mass, given the absence of observable gas in Leo II presently (though it is expected to be present at higher redshifts). The parameters $\chi_{\textit{star}}$, $\eta$, and $v_{\textit{wind,star}}$ used here were employed in the fiducial simulation outlined in the investigation of stellar feedback in Leo II \citep{Hazenfratz2024}. The mass of stellar particles obtained in the simulations was approximately $\sim2 \times 10^3$ M$_{\odot}$.   

\begin{table}
	\centering
	\caption{General configuration and parameters of the SPH simulations implementing AGN feedback for an isolated dwarf spheroidal galaxy, considering Leo II as a reference.}
	\label{tab:config_table}
	\begin{tabular}{lc} 
		\hline
		Parameter & Configuration/Value\\
		\hline
		$v_{200}$$^1$ & 20.5 km s$^{-1}$\\
		DM halo & Hernquist potential\\
		DM mass & $1.6 \times 10^9$ M$_{\odot}$\\
		DM concentration ($c$)$^2$ & 9\\
		Gas fraction ($m_b$) & 0.16\\
		Initial gas reservoir & $3.2 \times 10^8$ M$_{\odot}$\\
		Gas scale length ($b$) & 18.6 kpc\\
		Initial gas particle velocity & 0\\
		DM particle number  & 30000\\
		DM mass resolution & $5.3 \times 10^4$ M$_{\odot}$\\
		Gas particle number & 20000\\
	    Initial gas particle mass & $1.6 \times 10^4$ M$_{\odot}$\\
		Spin paramater ($\lambda$)$^3$ & 0.03\\
		Interpolation parameter ($f_{\text{eos}}$) & 1\\
		Gravitational softening length & adaptive\\
		Initial mass function & Chabrier\\
        $\chi_{\textit{star}}$ & 0.5\\
        Stellar mass loading factor $\eta$ & 60\\
        $v_{\textit{wind,star}}$ (km s$^{-1}$) & 96\\
		BH seed mass & $10^3 - 10^6$ M$_{\odot}$\\
        Radiative efficiency & 0.10 - 0.42\\
        Feedback efficiency & 0.01 - 0.05\\
        AGN wind geometry & isotropic/biconical\\
        \hline
        \end{tabular}
        \begin{minipage}{7cm}
        $^1$ \citet{Strigari2007}.\\
        $^2$ \citet{Correa2015, Cimatti2020}.\\
        $^3$ \citet{Bryan2013, Kurapati2018}.\\
        \end{minipage}
 \end{table}

For the investigation of AGN feedback in a dwarf spheroidal galaxy, we conducted simulations incorporating both thermal and thermal+kinetic feedback for each black hole seed mass ranging from $10^3$ to $10^6$ M$_{\odot}$. The selection of other parameters (refer to Table~\ref{tab:simul_table}) was guided by constraints on black hole masses observed in dwarf spheroidals \citep[e.g.,][]{Lora2009,Nucita2013,Manni2015,Lanfranchi2021}, as well as by previous cosmological simulations of massive galaxies \citep{Barai2014, Barai2019}. These simulations provided upper limits for dwarf AGN counterparts considered here.

\begin{table*}
	\centering
	\caption{Parameter space tested for the AGN feedback model in an isolated dwarf spheroidal galaxy.}
	\label{tab:simul_table}
        \resizebox{\textwidth}{!}{\begin{tabular}{cccccccccc} 
		\hline
	    Simulation & Label & M$_{\textit{BH,seed}}$ (M$_{\odot}$) & $\epsilon_{\textit{rad}}$ & $\alpha$ & Feedback type & $\epsilon_f$ & v$_w$ (km s$^{-1}$) & BH reposition & AGN wind geometry\\
		\hline
		1 & T3A10E5 & $10^3$ & 0.1 & 10 & thermal & 0.05 & - & no & isotropic\\
        2 & T3A50E1 & $10^3$ & 0.1 & 50 & thermal & 0.01 & - & no & isotropic\\
        3 & TK3A100E1V5 & $10^3$ & 0.1 & 100 & thermal+kinetic & 0.01 & 5000 & no & isotropic\\
        4 & T3A100E1 & $10^3$ & 0.1 & 100 & thermal & 0.01 & - & no & isotropic\\
        5 & T3A1000E1 & $10^3$ & 0.1 & 1000 & thermal & 0.01 & - & no & isotropic\\
        6 & T4A100E1 & $10^4$ & 0.1 & 100 & thermal & 0.01 & - & no & isotropic\\
        7 & TK4A100E1V5 & $10^4$ & 0.1 & 100 & thermal+kinetic & 0.01 & 5000 & no & isotropic\\
        8 & T5A100E1 & $10^5$ & 0.1 & 100 & thermal & 0.01 & - & no & isotropic\\
        9 & TK5A100E1V5 & $10^5$ & 0.1 & 100 & thermal+kinetic & 0.01 & 5000 & no & isotropic\\
        10 & TK3A100E1V5 & $10^3$ & 0.1 & 100 & thermal+kinetic & 0.01 & 5000 & no & isotropic\\
        11 & T6A100E1 & $10^6$ & 0.1 & 100 & thermal & 0.01 & - & no & isotropic\\
        12 & TK6A100E1V5 & $10^6$ & 0.1 & 100 & thermal+kinetic & 0.01 & 5000 & no & isotropic\\
        13 & TK3A100E1V5R & $10^3$ & 0.1 & 100 & thermal+kinetic & 0.01 & 5000 & yes & isotropic\\
        14 & TK4A100E1V5R & $10^4$ & 0.1 & 100 & thermal+kinetic & 0.01 & 5000 & yes & isotropic\\
        15 & TK5A100E1V5R & $10^5$ & 0.1 & 100 & thermal+kinetic & 0.01 & 5000 & yes & isotropic\\
        16 & TK6A100E1V5R & $10^6$ & 0.1 & 100 & thermal+kinetic & 0.01 & 5000 & yes & isotropic\\
        17 & TK6A1E1V5R & $10^6$ & 0.1 & 1 & thermal+kinetic & 0.01 & 5000 & yes & isotropic\\
        18 & T3A100E5 & $10^3$ & 0.1 & 100 & thermal & 0.05 & - & no & isotropic\\
        19 & TK3A100E5R & $10^3$ & 0.1 & 100 & thermal+kinetic & 0.05 & 5000 & yes & isotropic\\
        20 & T4A100E5 & $10^4$ & 0.1 & 100 & thermal & 0.05 & - & no & isotropic\\
        21 & TK4A100E5V5R & $10^4$ & 0.1 & 100 & thermal+kinetic & 0.05 & 5000 & yes & isotropic\\
        22 & T5A100E5 & $10^5$ & 0.1 & 100 & thermal & 0.05 & - & no & isotropic\\
        23 & TK5A100E5V5R & $10^5$ & 0.1 & 100 & thermal+kinetic & 0.05 & 5000 & yes & isotropic\\
        24 & T6A100E5 & $10^6$ & 0.1 & 100 & thermal & 0.05 & - & no & isotropic\\
        25 & TK6A100E5V5R & $10^6$ & 0.1 & 100 & thermal+kinetic & 0.05 & 5000 & yes & isotropic\\
        26 & TK3A100E5V3R & $10^3$ & 0.1 & 100 & thermal+kinetic & 0.05 & 3000 & yes & isotropic\\
        27 & TK4A100E5V3R & $10^4$ & 0.1 & 100 & thermal+kinetic & 0.05 & 3000 & yes & isotropic\\
        28 & TK5A100E5V3R & $10^5$ & 0.1 & 100 & thermal+kinetic & 0.05 & 3000 & yes & isotropic\\
        29 & TK6A100E5V3R & $10^6$ & 0.1 & 100 & thermal+kinetic & 0.05 & 3000 & yes & isotropic\\
        30 & TK3A100E5V1R & $10^3$ & 0.1 & 100 & thermal+kinetic & 0.05 & 1000 & yes & isotropic\\
        31 & TK4A100E5V1R & $10^4$ & 0.1 & 100 & thermal+kinetic & 0.05 & 1000 & yes & isotropic\\
        32 & TK5A100E5V1R & $10^5$ & 0.1 & 100 & thermal+kinetic & 0.05 & 1000 & yes & isotropic\\
        33 & TK6A100E5V1R & $10^6$ & 0.1 & 100 & thermal+kinetic & 0.05 & 1000 & yes & isotropic\\
        34 & TK3A100E1V3R & $10^3$ & 0.1 & 100 & thermal+kinetic & 0.01 & 3000 & yes & isotropic\\
        35 & TK4A100E1V3R & $10^4$ & 0.1 & 100 & thermal+kinetic & 0.01 & 3000 & yes & isotropic\\
        36 & TK5A100E1V3R & $10^5$ & 0.1 & 100 & thermal+kinetic & 0.01 & 3000 & yes & isotropic\\
        37 & TK6A100E1V3R & $10^6$ & 0.1 & 100 & thermal+kinetic & 0.01 & 3000 & yes & isotropic\\
        38 & TK3A100E1V1R & $10^3$ & 0.1 & 100 & thermal+kinetic & 0.01 & 1000 & yes & isotropic\\
        39 & TK4A100E1V1R & $10^4$ & 0.1 & 100 & thermal+kinetic & 0.01 & 1000 & yes & isotropic\\
        40 & TK5A100E1V1R & $10^5$ & 0.1 & 100 & thermal+kinetic & 0.01 & 1000 & yes & isotropic\\
        41 & TK6A100E1V1R & $10^6$ & 0.1 & 100 & thermal+kinetic & 0.01 & 1000 & yes & isotropic\\
        42 & K4A100E1V3R & $10^4$ & 0.1 & 100 & kinetic & 0.01 & 3000 & yes & isotropic\\
        43 & K5A100E1V3R & $10^5$ & 0.1 & 100 & kinetic & 0.01 & 3000 & yes & isotropic\\
        44 & TK4A10E1V3R & $10^4$ & 0.1 & 10 & thermal+kinetic & 0.01 & 3000 & yes & isotropic\\
        45 & TK5A10E1V3R & $10^5$ & 0.1 & 10 & thermal+kinetic & 0.01 & 3000 & yes & isotropic\\
        46 & TK4A1E1V3R & $10^4$ & 0.1 & 1 & thermal+kinetic & 0.01 & 3000 & yes & isotropic\\
        47 & TK5A1E1V3R & $10^5$ & 0.1 & 1 & thermal+kinetic & 0.01 & 3000 & yes & isotropic\\
        48 & TK4A100E1V3 & $10^4$ & 0.1 & 100 & thermal+kinetic & 0.01 & 3000 & no & isotropic\\
        49 & TK5A100E1V3 & $10^5$ & 0.1 & 100 & thermal+kinetic & 0.01 & 3000 & no & isotropic\\
        50 & T3A100E1R & $10^3$ & 0.1 & 100 & thermal & 0.01 & - & yes & isotropic\\
        51 & T4A100E1R & $10^4$ & 0.1 & 100 & thermal & 0.01 & - & yes & isotropic\\
        52 & T5A100E1R & $10^5$ & 0.1 & 100 & thermal & 0.01 & - & yes & isotropic\\
        53 & T6A100E1R & $10^6$ & 0.1 & 100 & thermal & 0.01 & - & yes & isotropic\\
        54 & TK4A100E1V3R-bi & $10^4$ & 0.1 & 100 & thermal+kinetic & 0.01 & 3000 & yes & biconical 30$^{\circ}$\\
        55 & TK5A100E1V3R-bi & $10^5$ & 0.1 & 100 & thermal+kinetic & 0.01 & 3000 & yes & biconical 30$^{\circ}$\\
        56 & TK4A100E1V3R-er42 & $10^4$ & 0.42 & 100 & thermal+kinetic & 0.01 & 3000 & yes & isotropic\\
        57 & TK5A100E1V3R-er42 & $10^5$ & 0.42 & 100 & thermal+kinetic & 0.01 & 3000 & yes & isotropic\\
        58 & TK3A100E1V3C & $10^3$ & 0.1 & 100 & thermal+kinetic & 0.01 & 3000 & center-fixed & isotropic\\
        59 & TK4A100E1V3C & $10^4$ & 0.1 & 100 & thermal+kinetic & 0.01 & 3000 & center-fixed & isotropic\\
        60 & TK5A100E1V3C & $10^5$ & 0.1 & 100 & thermal+kinetic & 0.01 & 3000 & center-fixed & isotropic\\
        61 & TK6A100E1V3C & $10^6$ & 0.1 & 100 & thermal+kinetic & 0.01 & 3000 & center-fixed & isotropic\\
		\hline
		\end{tabular}}
\end{table*}

Initially, we considered a range of feedback efficiency $\epsilon_f = [0.01, 0.25]$, based on studies involving more massive AGNs and dwarfs within a cosmological framework \citep{Barai2014, Bellovary2019}. However, our analysis focused on the narrower interval $\epsilon_f = [0.01,0.05]$ due to the significant impact of AGN feedback from IMBHs on galaxy evolution when $\epsilon_f = 0.05$ was adopted. In several simulations using this efficiency, star formation was excessively quenched. Consequently, AGN feedback efficiencies exceeding $\epsilon_f = 0.05$ were excluded from further consideration. The upper limit of $\epsilon_f = 0.05$ was also employed in studies of IMBHs through cosmological simulations by \citet{Barai2019} and in other works involving more massive AGNs, to bring the simulated $M_{BH}-\sigma$ relation in agreement to observations \citep[e.g.][]{DiMatteo2005, Springel2005}.

\section{Results and Discussion}

This section presents the analysis of results derived from the implementation and exploration of the parameter space of the AGN feedback model in a dwarf spheroidal galaxy. The primary focus is placed on understanding the growth and activity of black hole seeds within an isolated environment, as well as assessing the impact of this feedback on the evolution of our model galaxy.

\begin{figure*}
    \includegraphics[width=\textwidth]{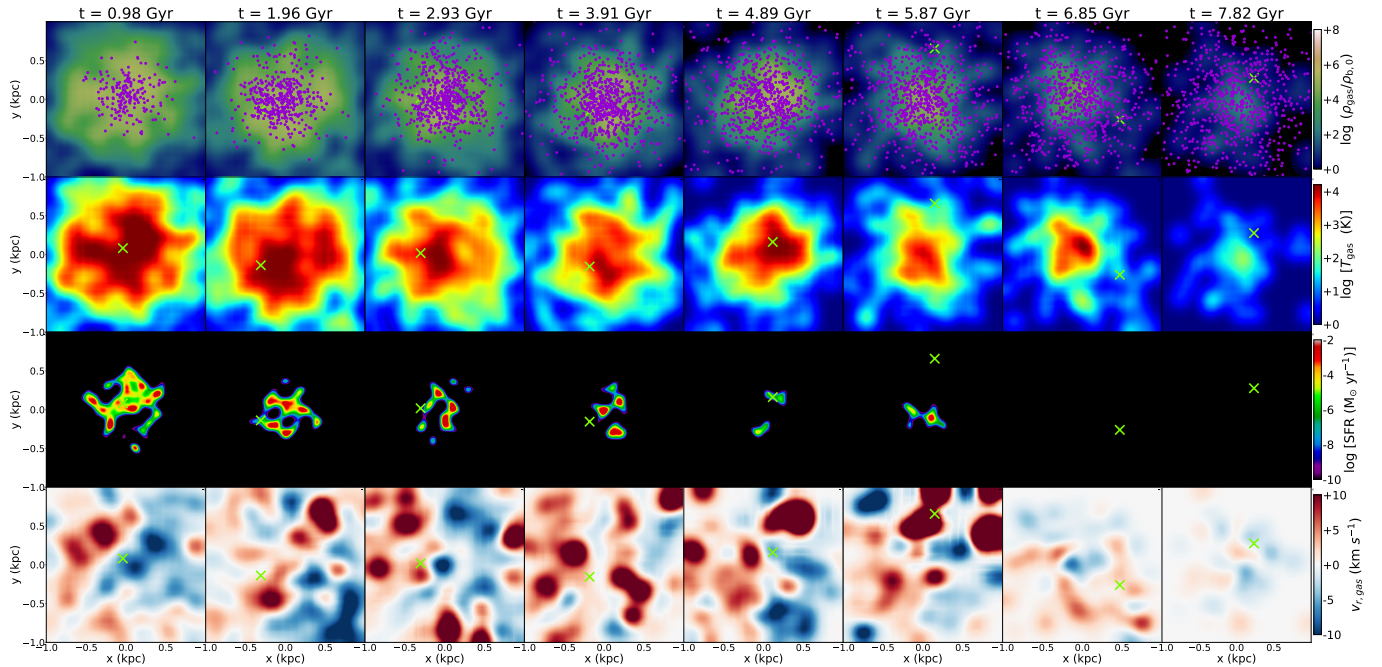}
    \vspace*{2mm}
    \caption{Maps in the xy-plane for gas overdensity (with star particle positions in magenta), temperature, star formation rate, and gas radial velocity at different times for simulation TK4A100E1V5R. The green cross indicates the BH position. The gas overdensity in the first row represents a contrast with the current mean baryon density of the universe (for $\Omega_{b,0} = 0.049$ - \citet[][]{Planck2018}). The fourth column depicts the radial component of gas velocity, where red indicates outward flows and blue represents inward flows.}
    \label{fig:gas_maps}
\end{figure*}

As an example, Fig.~\ref{fig:gas_maps} illustrates a 2D map for gas overdensity (a contrast with the current mean baryon density of the universe), temperature, star formation rate, and radial gas velocity (along the versor $\hat{r}$ in spherical coordinates) for a cut in the xy-plane in a chosen simulation. In these figures, the green cross denotes the instantaneous position of the black hole at the specified time snapshot of the simulation. Note that the black hole's position is not precisely at $(x,y) = (0,0)$ but instead exhibits a displacement within 1 kpc over time. Such wandering displacement tends to increase over time as the gas density in the central regions of the galaxy decreases.

\subsection{AGN kinematics and black hole growth}

The first question addressed by our simulations regarded the efficiency of BH seeds to grow within an isolated galaxy setting, where they would only be prone to growth by gas accretion, with no influence from mergers or environmental effects. Note that other growth mechanisms such as tidal disruption events (TDEs) are not included in the AGN models. 

\begin{figure*}
	\includegraphics[width=\textwidth]{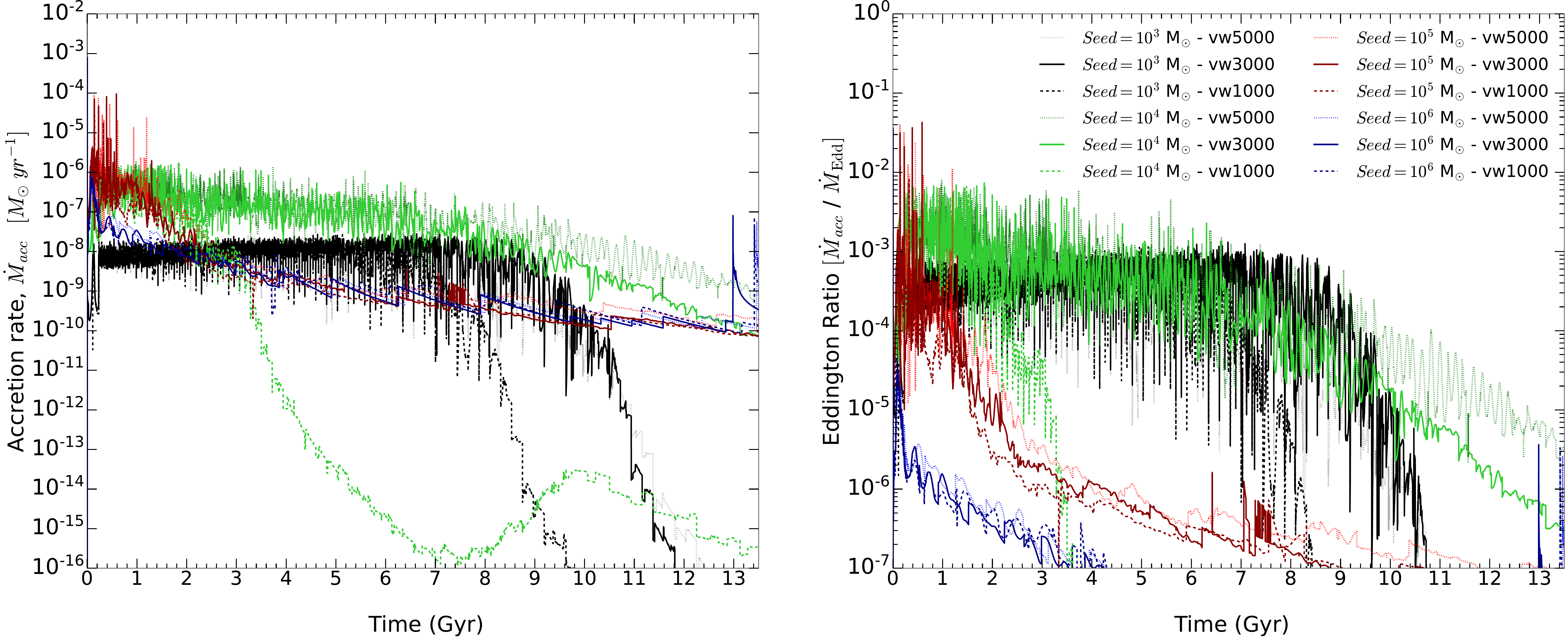}
    \vspace*{0mm}
    \caption{Black hole accretion rate and Eddington ratio ($\dot M_{\textit{BH}}/\dot M_{\textit{Edd}}$) for different black hole seeds and AGN wind velocities. Feedback efficiency: $\epsilon_\textit{f} = 0.05$.}
    \label{fig:BHER05}
\end{figure*}

\begin{figure*}
	\includegraphics[width=\textwidth]{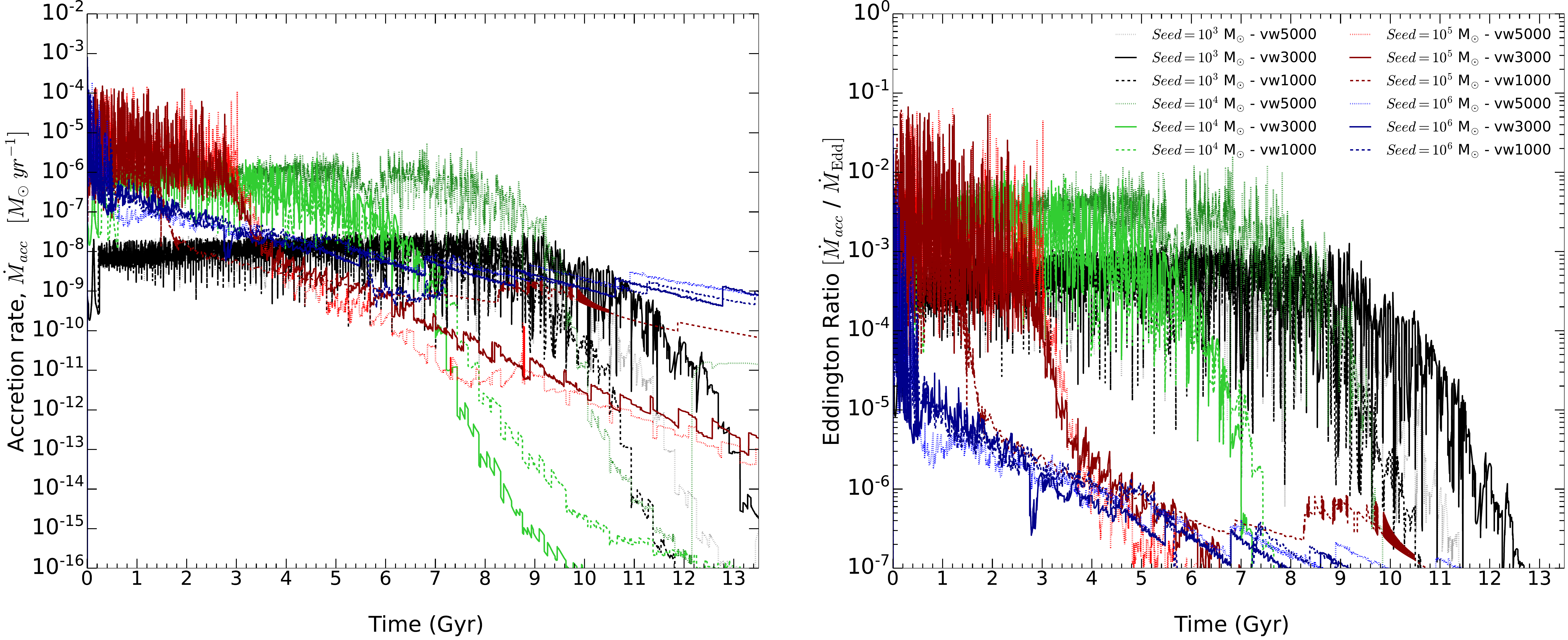}
    \vspace*{0mm}
    \caption{Black hole accretion rate and Eddington ratio ($\dot M_{\textit{BH}}/\dot M_{\textit{Edd}}$) for different black hole seeds and AGN wind velocities. Feedback efficiency: $\epsilon_\textit{f} = 0.01$.}
    \label{fig:BHER01}
\end{figure*}

The accretion rates and Eddington ratios ($\dot M_{\textit{BH}}/\dot M_{\textit{Edd}}$) calculated for selected simulations are displayed in Fig~\ref{fig:BHER05} for feedback efficiency of $5\%$ and in Fig~\ref{fig:BHER01} for $1\%$. In both graphs, there are considerable fluctuations between outbursts and more quiescent periods, according to what is expected for the small duty cycles of typical AGNs \citep[e.g.][]{Barai2014}. The most intense AGN activity occurs within the first 4 Gyr of galactic evolution in most cases. The exceptions are the simulations with $10^3$ M$_{\odot}$ BH seeds for both AGN feedback efficiencies, that present a nearly constant accretion activity until $\sim 9$ Gyr; and the simulation with $10^4$ M$_{\odot}$ and $v_w = 5000$ km$s^{-1}$, with a peak of activity after 5 Gyr, for $\epsilon_{f} = 0.01$. 

The intermittent regime of the gas accretion onto the BH is reproduced self-consistently and the Eddington ratios are generally lower than $1\%$ for $\epsilon_{\textit{feed}} = 0.05$ and $7\%$ for $\epsilon_{\textit{feed}} = 0.01$. The higher Eddington ratios for the lower AGN feedback efficiency are related mainly to the higher availability of gas when that parameter is reduced, due to the diminished gas outflow rate from kinetic feedback (Eq.~\ref{eq:outflow-rate}). 

The alternating periods of higher and lower AGN activity typically endure for less than 50 Myr. The decline in this activity over time can be attributed to a decrease in gas density resulting from the combined effects of stellar and AGN feedback in the central region of the galaxy, where most of the star formation and accretion activity happens in the simulations. 

The accretion rates onto the IMBHs depicted in Fig~\ref{fig:BHER05} were observed within the interval of approximately $10^{-8}$ to $10^{-4}$ M$_{\odot}$ yr$^{-1}$ over the first 4 Gyr of galactic evolution, for the case with an AGN feedback efficiency of $\epsilon_{\text{feed}} = 0.05$, with the highest values recorded for the $10^4$ and $10^5$ BH seeds. A similar rate interval was observed in Fig~\ref{fig:BHER01} for $\epsilon_{\text{feed}} = 0.01$, although with higher values compared to the $\epsilon_{\text{feed}} = 0.05$ case. The highest accretion rates were noted for $10^5$ and $10^6$ BH seeds.

According to \citet{Barai2019}, who conducted cosmological simulations for AGNs in dwarf galaxies, accretion rates ranging approximately from $10^{-7}$ to $10^{-2}$ M$_{\odot}$ yr$^{-1}$ were found for the most massive IMBH in each run for final redshifts $z = [4-7.5]$, corresponding roughly to universe ages between 0.7 and 1.6 Gyr. Their rates can reach values approximately three orders of magnitude higher than the maximum accretion rates found in our study. Moreover, they found the BH accretion rates increase with time and reach $\dot M_{\textit{BH}} \sim (0.2-0.8)\dot M_{Edd}$ for the massive IMBHs by z = 4. This comparison might indicate a reduced accretion activity within an isolated environment, when compared to the black hole growth in galaxies prone to mergers and tidal interactions.

\subsubsection{Influence of the BH seed mass}

\begin{figure*}
	\includegraphics[width=\textwidth]{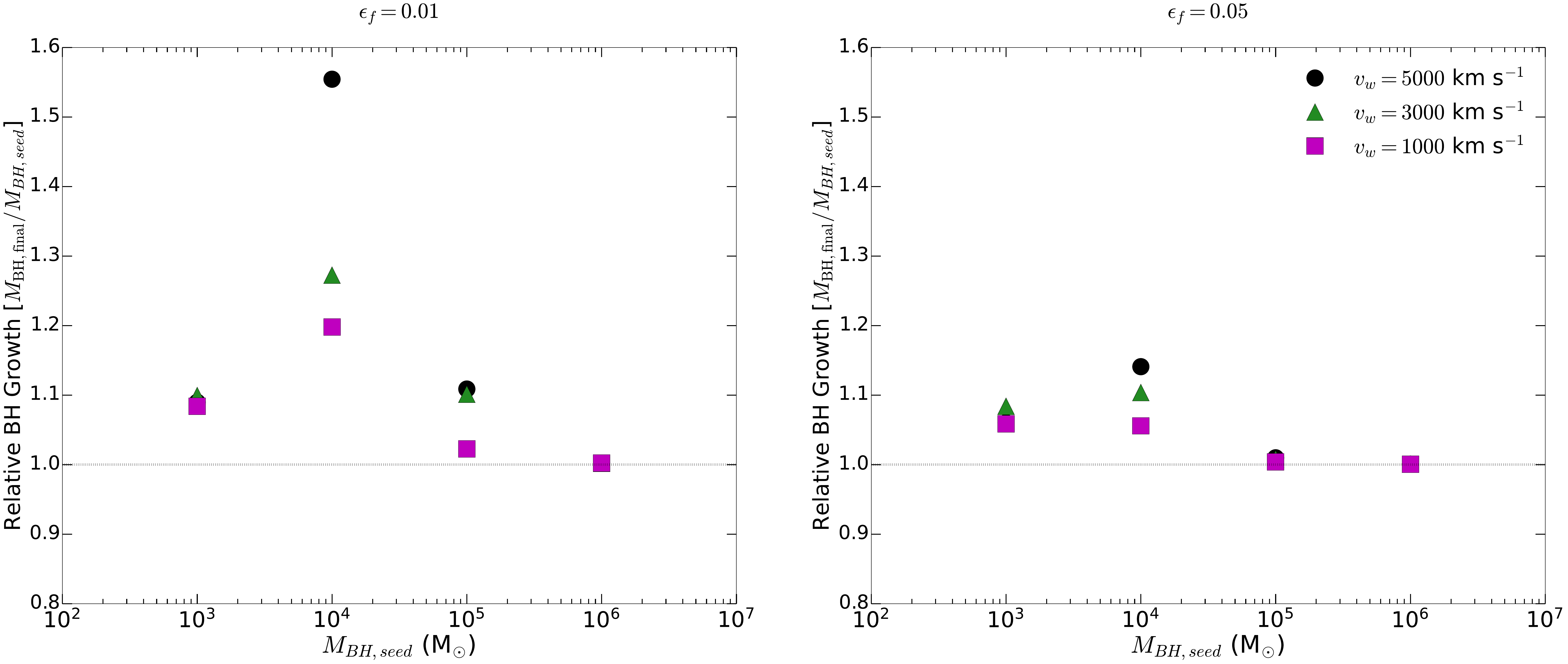}
    \vspace*{0mm}
    \caption{Influence of black hole seed mass and AGN wind injection velocity on black hole growth.}
    \label{fig:BH-growth}
\end{figure*}

In general, the BH seeds with $10^5$ M$_{\odot}$ produced the highest accretion rates and Eddington ratios in Figs~\ref{fig:BHER05} and \ref{fig:BHER01}, for the first 3 Gyr of galactic evolution. To analyze the integrated effect of BH seed mass on the IMBH growth, the relative growth ($M_{\textit{BH, final}}$/$M_{\textit{seed}}$) of the seeds were calculated for each case. The results are displayed in Fig.~\ref{fig:BH-growth}.

Fig.~\ref{fig:BH-growth} shows a limited growth for the black hole irrespective of the BH seed mass, AGN feedback efficiency and AGN wind velocity. The highest relative growth was observed for BH seeds of $10^4$ M$_{\odot}$, suggesting an optimal balance between accretion and feedback for the BH growth from this seed mass. The ratios $M_{\textit{BH, final}}$/$M_{\textit{seed}}$ consistently remained below 2, indicating a marginal growth within the isolated environment of our simulations, as the order of magnitude of the seed masses remained unchanged. This, coupled with the potential for even lower feedback efficiencies for dwarf AGNs (we started considering upper limits for the model´s free parameters, derived from AGN simulations with disk galaxies) suggests a significant role played by mergers and interactions in the growth of IMBHs in dwarf galaxies.

\subsubsection{Influence of feedback efficiency}

We explored two feedback efficiencies in Fig.~\ref{fig:BH-growth}. A more pronounced BH growth was observed for $\epsilon_f = 0.01$ compared to $\epsilon_f = 0.05$, although the difference was limited to $\sim 40 \%$ in  $M_{\textit{BH, final}}$/$M_{\textit{seed}}$. This difference can be attributed to the higher AGN feedback efficiency resulting in higher gas outflow rates, consequently depriving the galactic central regions of gas and further star formation. This observation aligns with findings from \citet{Hazenfratz2024}, where the region within the tidal radius was identified as the primary area containing all the star-forming particles in most of the simulations focusing on stellar feedback in Leo II.

\subsubsection{Influence of AGN wind injection velocity}

In Fig.~\ref{fig:BH-growth}, we observed that the BH growth increases with higher AGN wind velocities in general, particularly evident for a BH seed mass of $10^4$ M$_{\odot}$, with a reduced influence on other seeds. This trend can be understood by considering Eq.~\ref{eq:outflow-rate}, where increasing the value of the AGN wind velocity $v_{w}$ results in a decrease in the gas outflow rate $\dot M_w$ ($\dot M_w \propto v_{w}^{-2}$), assuming $\dot M_{\textit{BH}}$ remains relatively constant (which is observed approximately for the first 2 Gyr in Figs \ref{fig:BHER05} and \ref{fig:BHER01}). This would lead to a greater availability of gas around the BH. Note, however, that for $10^3$ M$_{\odot}$, slightly higher growth rates were actually observed for the intermediate AGN wind velocity of $v_{w} = 3000$ km $s^{-1}$.

\subsection{Galactic evolution} \label{SFH}

This section focuses on the analysis of the AGN feedback impact on the galactic evolution of our benchmark galaxy. For comparison purposes, we have selected a fiducial isolated simulation, with stellar-only feedback, from \citet{Hazenfratz2024}, which provided the best overall fit for the constraints chosen to model a dwarf-like galaxy with characteristics similar to Leo II. In this regard, the simulation $\eta$60vCalc was chosen as the fiducial simulation to analyze the AGN influence. It features a fraction $\chi = 0.5$ for the supernovae energy coupled to the wind kinetic energy, mass loading factor $\eta = 60$ and wind kick velocity $v_{\textit{wind}} \sim96$ km s$^{-1}$ for the stellar feedback model. 

\subsubsection{Star formation history}

\begin{figure*}
	\includegraphics[width=\textwidth]{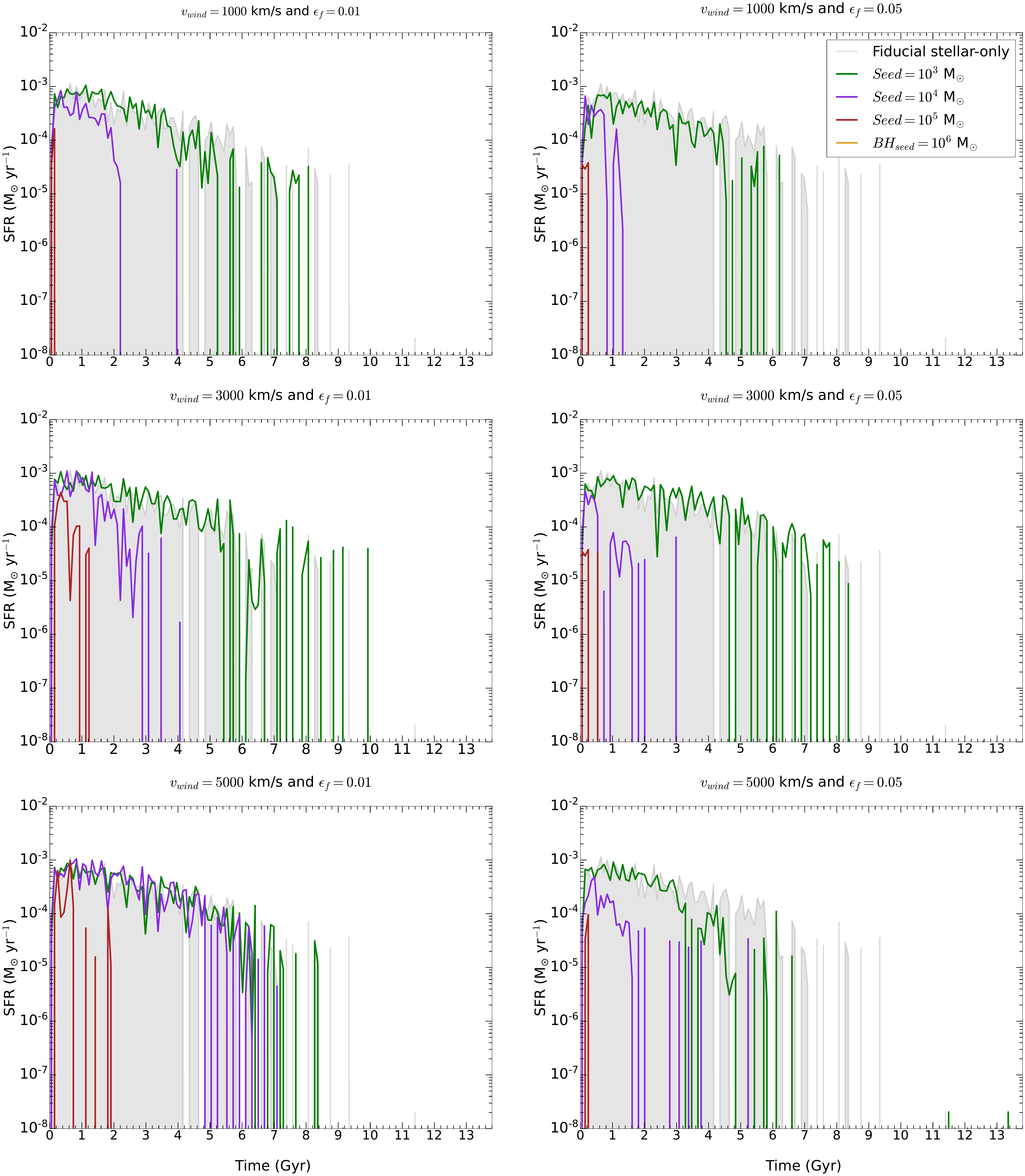}
    \vspace*{0mm}
    \caption{Influence of AGN feedback on the star formation history of a simulated dwarf spheroidal galaxy, comprising different black hole seed masses and AGN wind injection velocities. The gray-shaded curve represents the fiducial simulation of Leo II with stellar feedback only, from \citet{Hazenfratz2024}.}
    \label{fig:sfr-combined}
\end{figure*}

We start our analysis of the dwarf AGN feedback by evaluating its impact on the star formation history (SFH) of the simulated galaxies. In Figure \ref{fig:sfr-combined}, the SFH of the fiducial simulation with stellar-only feedback is juxtaposed with simulations incorporating AGN feedback. The yellow curve corresponding to $10^6$ M$_{\odot}$ does not appear because star formation is completely suppressed in simulations with a black hole of this mass. A general comparison of the plots for different AGN feedback efficiencies show that the star formation quenching was more pronounced for $\epsilon_f = 0.05$. As discussed earlier, this trend can be attributed to a reduced availability of gas for star formation within the central region of the galaxy, where all the SF occurred in our simulations. 

\begin{figure*}
	\includegraphics[width=\textwidth]{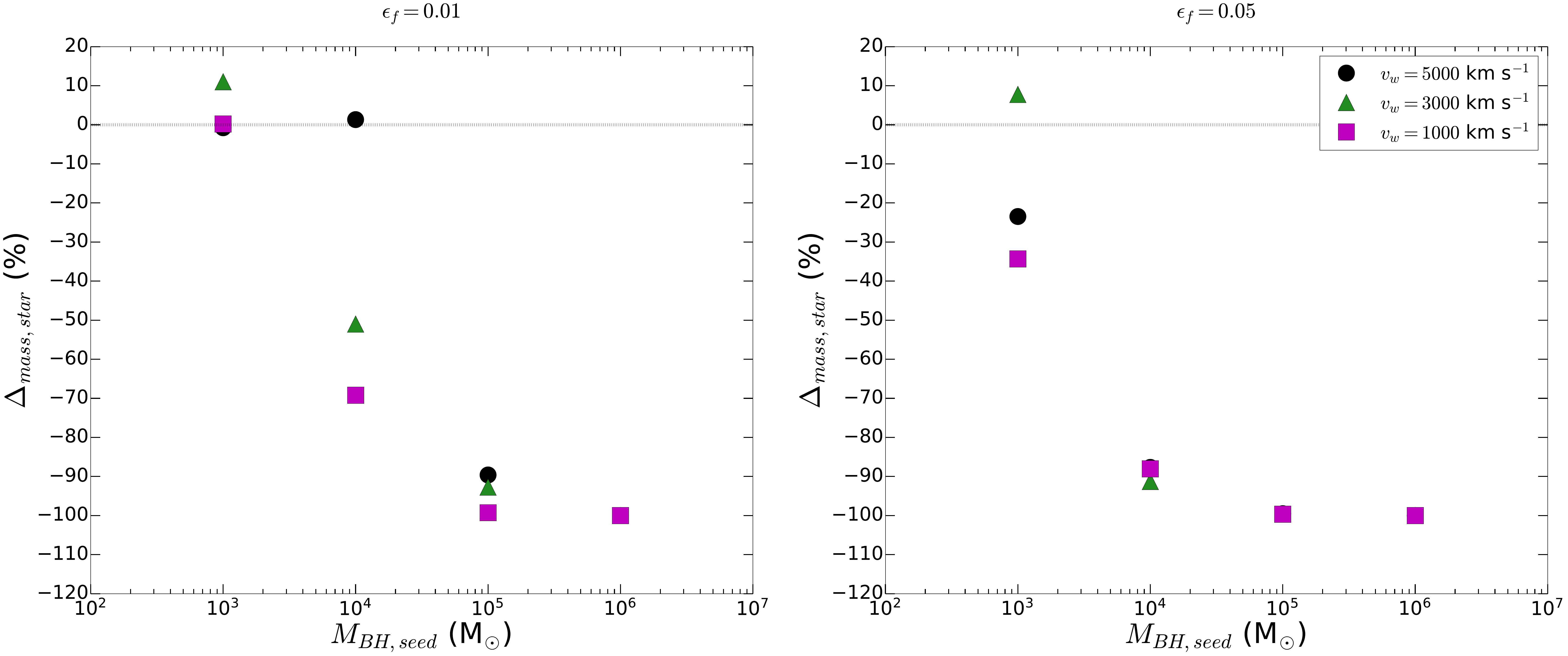}
    \vspace*{0mm}
    \caption{Influence of AGN feedback on the final stellar mass of a simulated dwarf spheroidal galaxy, evaluated as the percentage difference compared to the stellar mass generated in the fiducial simulation with stellar feedback only, from \citet{Hazenfratz2024}.}
    \label{fig:delta-star}
\end{figure*}

When comparing different AGN wind velocities, two different trends were observed, depending on the black hole seed mass. For seeds of $10^3$ M$_{\odot}$, AGN wind velocities of $v_w = 1000$ and $v_w = 5000$ km s$^{-1}$ exhibited similar trends with a discrete reduction in star formation activity for $\epsilon _f = 0.01$, particularly evident when the star formation becomes more episodic, after 5 Gyr of galactic evolution. Conversely, for $v_w = 3000$ km s$^{-1}$, there was a discrete increase in the SF activity, resulting in higher final stellar masses, which are depicted in Fig.~\ref{fig:delta-star} for other BH seed masses as well. 

Fig.~\ref{fig:delta-star} confirms the impact of BH seeds of $10^3$ M$_{\odot}$, wherein AGN wind velocities of $v_w = 1000$ and $v_w = 5000$ km s$^{-1}$ resulted in practically identical final stellar masses. An increase of $\sim 11\%$ was observed for $v_w = 3000$ km s$^{-1}$ and $\epsilon_f = 0.01$, indicative of positive AGN feedback. For $\epsilon_f = 0.05$, a similar general trend was observed for seeds of $10^3$ M$_{\odot}$, but with a reduction in final stellar mass for $v_w = 1000$ (-34\%) and $v_w = 5000$ (-23\%) km s$^{-1}$. The increase in final stellar mass for $v_w = 3000$ km s$^{-1}$ was $\sim 8\%$ in this case.

The positive AGN feedback observed for $10^3$ M$_{\odot}$, and AGN wind velocities of $v_w = 3000$ km s$^{-1}$ can be correlated with the lower accretion rates for these seed masses (see Fig.~\ref{fig:BHER01} and Fig.~\ref{fig:BHER05}). These rates are not enough to deprive the galaxy of gas prone to star formation, but may facilitate shocks capable of compressing gas in the central regions of the galaxy. The fact that it occurred only for $v_w = 3000$ km s$^{-1}$ suggests that this intermediate value for the AGN wind velocity struck an optimal balance with the gas outflow rate, thereby creating the ideal amount of turbulence to enhance star formation. 

Different trends were observed for the other BH seed masses. The impact on the stellar formation activity for $10^4$ M$_{\odot}$ in Fig.~\ref{fig:sfr-combined} increases with decreasing AGN wind velocity for both feedback efficiencies, but leading to more pronounced differences among them for $\epsilon_f = 0.01$. For $v_w = 5000$ km s$^{-1}$ and $\epsilon_f = 0.01$, the impact in the star formation was minimum and a slight increase in the final stellar mass of $\sim 1.3\%$ was observed in Fig.~\ref{fig:delta-star} (which actually can also be attributed to stochastic fluctuations in star formation), while the other velocities resulted in significant negative feedback, reducing the final stellar mass by approximately $-50\%$ for $v_w = 3000$ km s$^{-1}$ and $-70\%$ for $v_w = 1000$ km s$^{-1}$. Conversely, for $\epsilon_f = 0.05$, the impact on the final stellar mass was comparable across the three AGN wind velocities, with stellar mass variations ranging between -88$\%$ and -91$\%$. This decreased stellar formation for these cases may indicate that the choice of $\epsilon_f = 0.05$ is excessively high for a dwarf spheroidal galaxy similar to Leo II.

The BH seeds of $10^5$ and $10^6$ M$_{\odot}$ exerted a critical influence on the evolution of the dwarf galaxy, as evidenced by the SFHs depicted in Fig.~\ref{fig:sfr-combined} and the stellar masses variations in Fig.~\ref{fig:delta-star}. Final stellar mass reductions were observed to range between $-90\%$ and $-99\%$ for $M_{\textit{BH,seed}}=10^5$ M$_{\odot}$. Moreover, no star formation activity was detected for $M_{\textit{BH,seed}}=10^6$ M$_{\odot}$ across all combinations of $\epsilon_f$ and $v_w$. Similar to the case with $M_{\textit{BH,seed}}=10^4$, these results suggest that the choice of $\epsilon_f = 0.05$, while plausible for AGNs in larger galaxies \citep[e.g.][]{DiMatteo2005, Springel2005,Barai2019}, may be excessively high for the lower end of the galactic mass function. The detection of potential signatures in X-rays and radio from IMBHs with masses ranging between $10^4-10^5$ M$_{\odot}$ in dwarf galaxies with masses similar to Leo II´s mass ($\sim10^9$ M$_{\odot}$) supports this conjecture \citep[e.g.][]{Maccarone2005,Lora2009,Nucita2013}. 

\begin{figure*}
	\includegraphics[width=\textwidth]{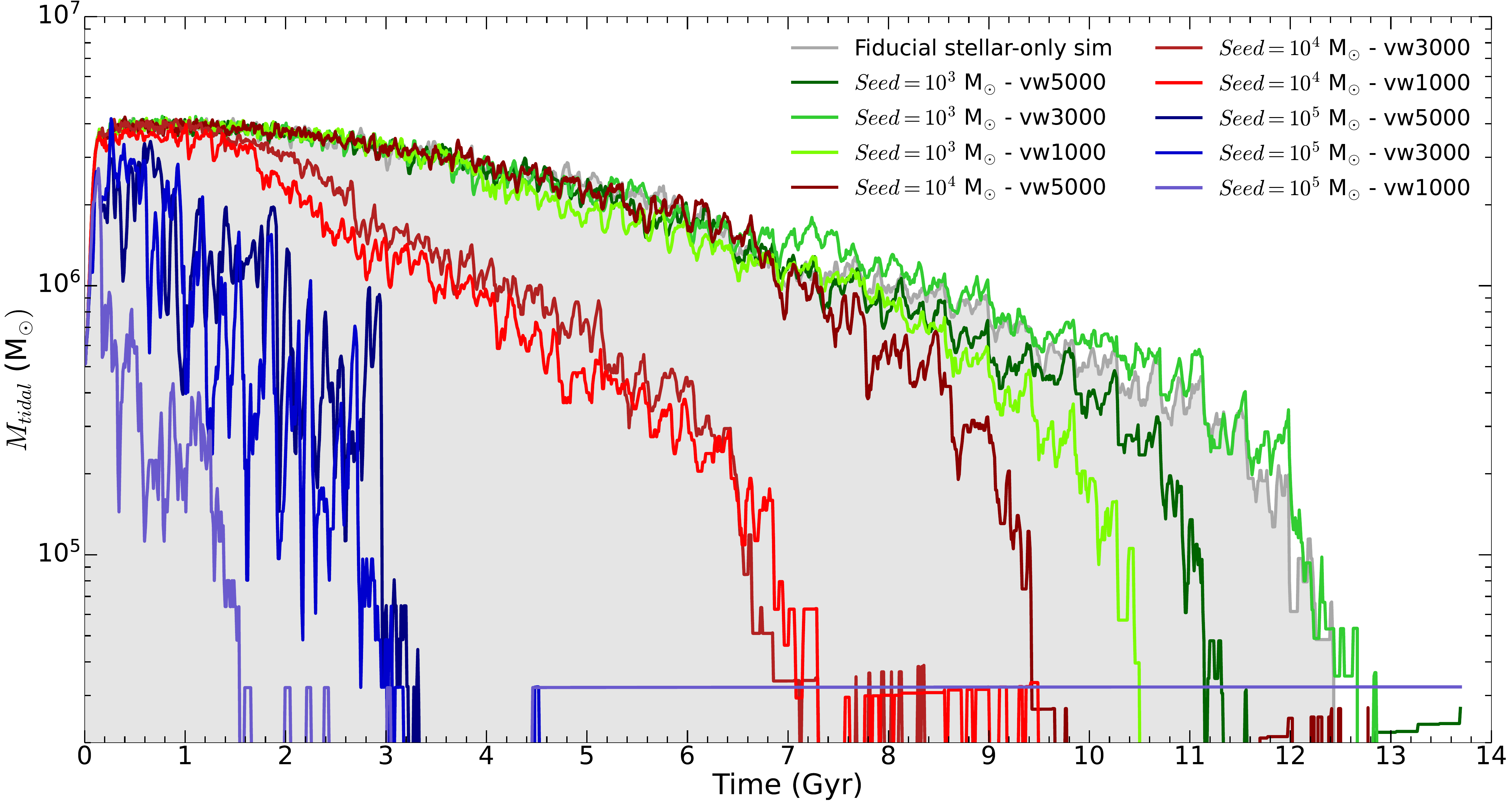}
    \vspace*{0mm}
    \caption{Influence of AGN feedback on the gas depletion within the tidal radius ($\sim 650$ pc) of a simulated dwarf spheroidal galaxy for $\epsilon_f = 0.01$.}
    \label{fig:mass-tidal}
\end{figure*}

One of the main characteristics to be reproduced in simulations of dwarf spheroidal galaxies is the exhaustion of their gaseous content as observed today, as these systems are often found to be devoid of gas \citep{Mateo1998, Strigari2007, Grcevich2009, Mcconnachie2012}. To assess the influence of a putative IMBH on the gas evolution of our fiducial dwarf galaxy, the gas content within the tidal radius ($\sim 650$ pc) is depicted in Fig.~\ref{fig:mass-tidal}. BH seeds of $10^6$ M$_{\odot}$ were not analyzed because they prevented any star formation in the galaxy.

The fiducial stellar-only simulation starts with an increase in gas mass until $\sim200$ Myr, reaching a peak of $\sim4 \times 10^6$ M$_{\odot}$ (Fig.~\ref{fig:mass-tidal}). This increase is attributed to gas falling towards the central region of the galaxy due to gravity and correlates with the rise in star formation rate up to its peak at $\sim 10^{-3}$ M$_{\odot}$ yr$^{-1}$ around 600 Gyr in Fig.~\ref{fig:sfr-combined}. During this period, the amount of thermal and kinetic energy released by stars into the gas also increases. Consequently, from $\sim 1$ Gyr onwards, there is a continuous depletion of gas until $\sim 12.4$ Gyr, when the tidal region becomes finally devoid of gas. The abrupt decline in gas mass after 12 Gyr is related to the point at which the mass of the residual gas becomes comparable to the mass of individual gas particles in our simulations ($\sim10^4$ M$_{\odot}$). Once the last particle outflows, the gas mass within the considered volume drops to zero. 

When a black hole is present, notable differences in gas dynamics emerge. For BH seeds of $10^3$ M$_{\odot}$, the amount of gas over time within the tidal radius is practically the same until $\sim 4$ Gyr, when comparing simulations with and without AGN. Note that this period corresponds approximately to the phase when the SFH with and without AGNs exhibit the greatest similarity in Fig.~\ref{fig:sfr-combined}, with more variation thereafter. Subsequently, differences in the curves arise depending on the AGN wind velocity. Specifically, for $v_w = 1000$ and $v_w = 5000$ km s$^{-1}$, gas depletion accelerates, being practically completed around 10.5 Gyr for $v_w = 1000$ km s$^{-1}$ and 11.6 Gyr for $v_w = 5000$ km s$^{-1}$. Conversely, for $v_w = 3000$ km s$^{-1}$, the gas depletion is delayed by $\sim 400$ Myr, concluding at $t = 12.8$ Gyr. Moreover, between $t \sim [6.2-12.8]$ Gyr, there are periods where the gas quantity within the tidal radius surpasses that observed in the fiducial simulation. Note that this AGN wind velocity also corresponds to the case exhibiting positive AGN feedback in Fig.~\ref{fig:delta-star}.

For BH seeds of $10^4$ M$_{\odot}$, the gas dynamics remain largely consistent across different AGN wind velocities until $\sim$ 1 Gyr. In all cases, the gas depletion is accelerated, with the slowest depletion observed for $v_w = 5000$ km s$^{-1}$, although one or two particles may have persisted for some time. Note that there is gas within the tidal radius for all AGN wind velocities until $\sim 7$ Gyr. This availability of gas, even after star formation has ceased, contrasts with the stellar formation quenching observed for these seeds in Fig.~\ref{fig:sfr-combined}, particularly notable for $v_w = 1000$ and $v_w = 3000$ km s$^{-1}$. This trend indicates the efficiency of AGN feedback in suppressing further star formation in the central region of the galaxy when a IMBH of $10^4$ M$_{\odot}$ is present under the chosen physical conditions. 

Similarly, for BH seeds of $10^5$ M$_{\odot}$, the gas depletion is accelerated accross all the AGN wind velocities, culminating in total gas depletion by $\sim 3.4$ Gyr, although one or two gas particles could appear in later times. In this scenario, the galaxy is deprived of gas faster, effectively halting any further star formation after 2 Gyr. Note that for the fiducial simulation with stellar-only feedback, the initial and continuous period of star formation persists until $\sim 4.2$ Gyr. However, this phase does not occur for the BH seeds of $10^5$ M$_{\odot}$ in Fig.~\ref{fig:sfr-combined}, indicating the efficiency of the AGN feedback in critically quenching the star formation under the selected physical conditions. 

The mass variations of gas within the tidal radius of the simulated galaxy, as depicted in Fig.~\ref{fig:mass-tidal}, do not follow a strictly monotonic pattern. Instead, they exhibit alternating periods of inflow and outflow. This behavior can be attributed to the continuous input of energy from stellar and AGN feedback mechanisms over time, which serves to counterbalance the gravitational pull on the gas reservoir. Furthermore, the oscillations in gas mass observed during the last 4 Gyr of galactic evolution remain consistent with the upper limit of $10^4$ M$_{\odot}$ for the HI mass in Leo II, as estimated by \citet{Grcevich2009}.

\subsubsection{Indirect AGN influence}

In addition to direct influences on observational constraints for dwarf galaxies, there exists a possibility of indirect effects of dwarf AGNs on galactic evolution. For example, AGN feedback could, in principle, change the location of supernovae over time due to perturbation in the gas dynamics and affect the evolution of the galaxy by altering the location of stellar feedback input, even if the influence on the star formation history is negligible. To evaluate this type of effect in our simulations, we examined the final distribution of stars in the simulated galaxy by estimating local stellar densities and the global quantity density radius. This analysis followed the method outlined by \citet{VonHoerner1963} for N-body systems, which was generalized by \citet{Casertano1985}. In this algorithm, the density around a star, adapted to varying particle masses, is defined by

\begin{equation}
    \label{eq:star_dens}
    \rho_j^{(i)} = \left(\sum\limits_{k=1}^{j-1}m_{*k} \right) / {V(r_j)}
\end{equation}

where $\rho_j^{(i)}$ is the density estimator of order \textit{j} around the $i$th star particle; $m_{*k}$ is the mass of the $k$th nearest star particle neighbor around $i$; $V(r_j)$ is the spherical volume determined by the distance $r_j$ to the $j$th nearest neighbor. \citet{Casertano1985} demonstrated that employing the order $j=6$, as chosen in our analysis, offers a convenient compromise to minimize local fluctuations while retaining the locality of the estimator. 

The density radius was then defined by \citet{VonHoerner1963} as

\begin{equation}
    \label{eq:dens_radius}
    r_{d,j} = \frac{\sum_{i} |\bf{x_{\textit{i}}} - \bf{x_{\textit{d,j}}}|\rho_{\textit{j}}^{(\textit{i})}}{\sum_{i} \rho_j^{(i)}}
\end{equation}

where the numerator of Eq.~\ref{eq:dens_radius} comprises the density-weighted average of the distances of each star particle from the density center, defined by

\begin{equation}
    \label{eq:dens_center}
    \bf{x_{\textit{d,j}}} = \frac{\sum_{i} \bf{x_{\textit{i}}} \rho_{\textit{j}}^{(\textit{i})}}{\sum_{i} \rho_{\textit{j}}^{(\textit{i})}}
\end{equation}

which represents a density-weighted average of the positions of star particles. 

\begin{figure*}
	\includegraphics[width=\textwidth]{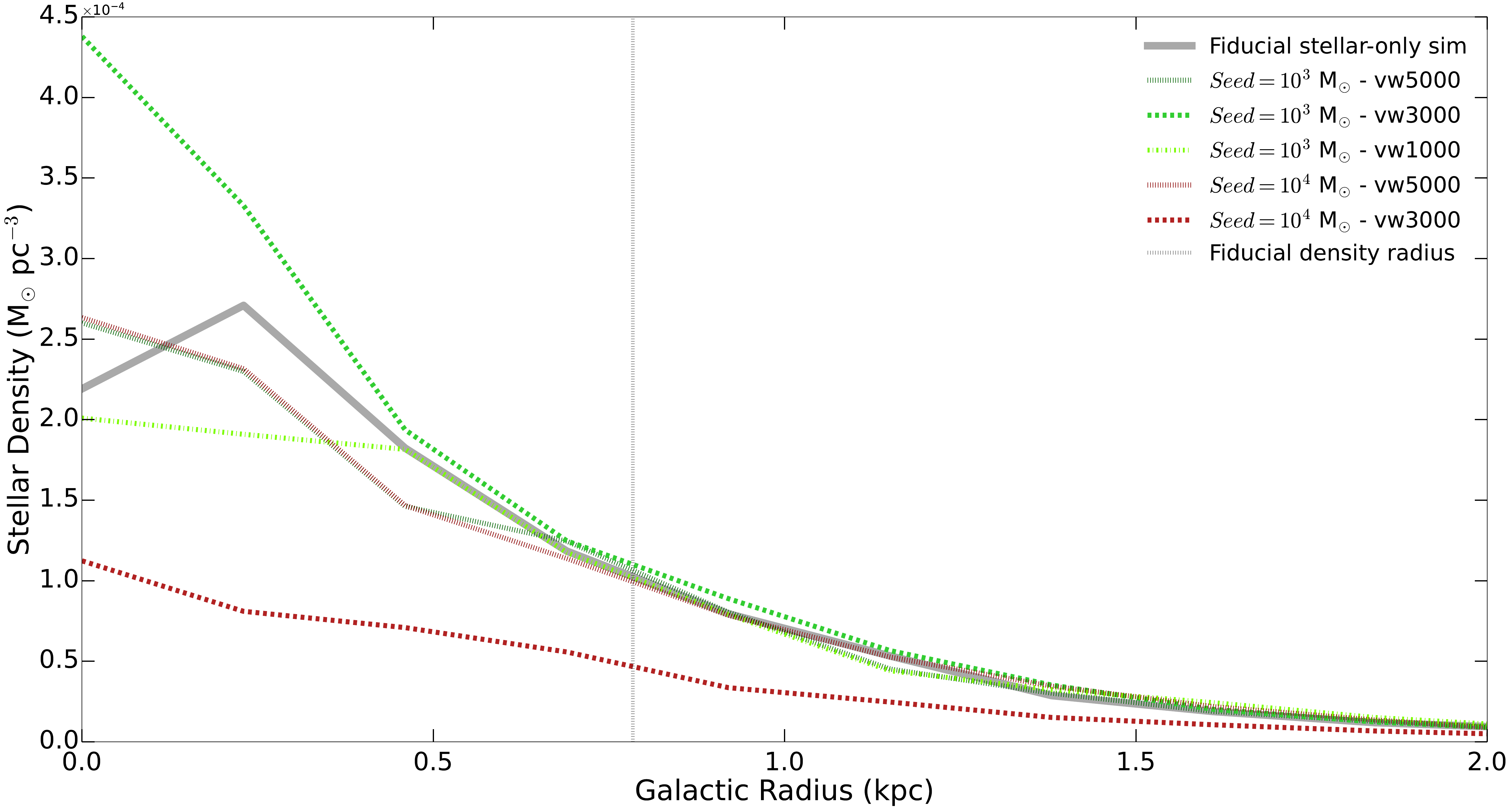}
    \vspace*{0mm}
    \caption{Stellar density profiles for selected simulated galaxies with $\epsilon_f = 0.01$.}
    \label{fig:star-dens}
\end{figure*}

For the subsequent analysis, we selected a subset of simulations characterized by the lowest discrepancies in final stellar mass. This approach enabled us to isolate and evaluate any potential indirect influence of the AGN on the galaxy's evolution. The chosen simulations include TK3A100E1V5R, TK3A100E1V3R, TK3A100E1V1R, and TK4A100E1V5R, all exhibiting final stellar mass variations below 50$\%$ as depicted in Fig.~\ref{fig:delta-star}. In Fig.~\ref{fig:star-dens}, we present the radial profiles of the stellar density estimator for these selected simulations at 13.7 Gyr. 

The majority of the profiles in Fig.~\ref{fig:star-dens} show similar behavior after the density radius estimated for the fiducial simulation with stellar-only feedback ($r_{d,j} = 784$ pc), except for the case with $M_{\textit{BH,seed}} = 10^4$ M$_{\odot}$ and $v_w = 3000$ km s$^{-1}$, where the curve for the stellar density profile consistently lies below the others. However, this difference can be more attributable to the reduction in final stellar mass (see Fig.~\ref{fig:delta-star}). The differences in the remaining profiles appear within the density radius, with lower values of stellar densities observed across all cases compared to the fiducial simulation without AGN, except for $M_{\textit{BH,seed}} = 10^3$ M$_{\odot}$ and $v_w = 3000$ km s$^{-1}$ and an inversion of behavior at the centre for $M_{\textit{BH,seed}} = 10^3, 10^4$ M$_{\odot}$ and $v_w = 5000$ km s$^{-1}$. 

\begin{figure*}
	\includegraphics[width=\textwidth]{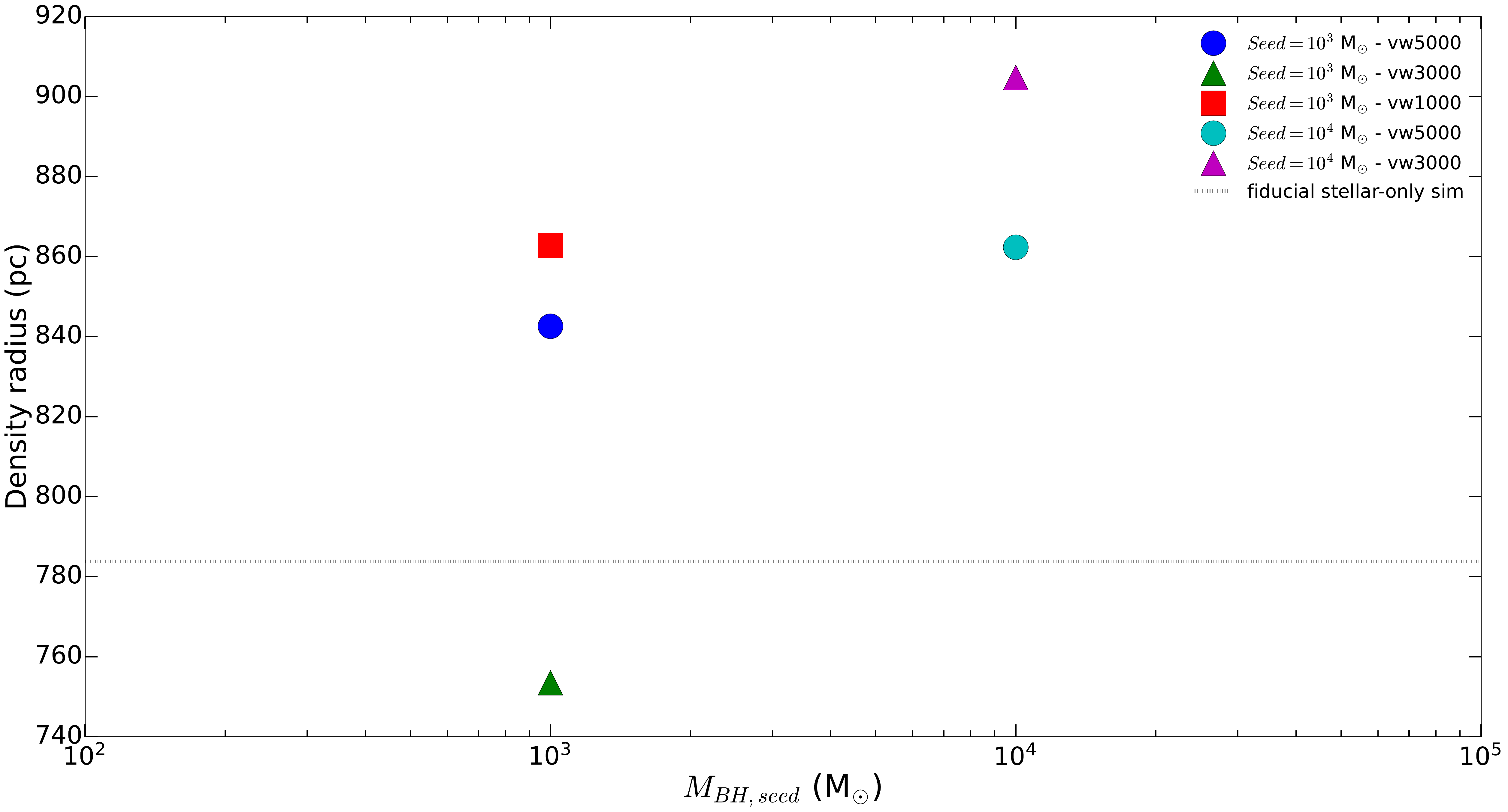}
    \vspace*{0mm}
    \caption{Stellar density radius estimated for selected simulated galaxies with $\epsilon_f = 0.01$.}
    \label{fig:dens-radius}
\end{figure*}

To further evaluate potential influences on the stellar distribution of simulated galaxies, we examined the density radius estimators, as shown in Fig.~\ref{fig:dens-radius}. It was observed that all simulations exhibited higher values for the density radius estimate, with the exception of the simulation featuring $M_{\textit{BH,seed}} = 10^3$ M$_{\odot}$ and $v_w = 3000$ km s$^{-1}$. This particular simulation displayed a decrease of $\sim 4\%$ in this measure. Notably, as depicted in Fig.~\ref{fig:delta-star}, this simulation also produced a positive AGN feedback, resulting in an increase of approximately 10$\%$ in the final stellar mass. It is plausible that this additional stellar mass contributed to a slightly more concentrated stellar distribution, owing to increased gravitational attraction towards the central regions of the galaxy. Additionally, as illustrated in Fig.~\ref{fig:star-dens}, the stellar density values across galactic radius were higher than those in the fiducial simulation until approximately 2 kpc.

The remaining four simulations exhibited higher variations in stellar density radius. For $M_{\textit{BH,seed}} = 10^3$ M$_{\odot}$ and $v_w = 5000$ km s$^{-1}$, the variation was $\sim 8\%$, and for $M_{\textit{BH,seed}} = 10^3$ M$_{\odot}$ and $v_w = 1000$ km s$^{-1}$, the difference was $\sim 10\%$ in the estimators, indicating more dispersed stellar distributions. These two simulations have in common the negligible difference in the final stellar mass in Fig.~\ref{fig:delta-star}. Additionally, as depicted in Fig.~\ref{fig:mass-tidal}, these two simulations were the ones that exhibited accelerated gas depletion for these BH seeds within the observationally estimated tidal radius of Leo II. This trend can then be attributed more to an altered spatial distribution of the stellar feedback in the galaxy, considering the reduced impact of AGN feedback for these BH seeds (see Fig.~\ref{fig:lum_kin}). Therefore, it can be inferred that the differences in the stellar distribution for these simulations may be linked to the presence of an IMBH of $10^3$ M$_{\odot}$ in this galaxy. 

In the simulations with $M_{\textit{BH,seed}} = 10^4$ M$_{\odot}$, the one with $v_w = 3000$ km s$^{-1}$ had an increment of $\sim 15\%$ in this quantity. This simulation also presented a reduction of $\sim 50\%$ in the final stellar mass (Fig.~\ref{fig:delta-star}). Consequently, the increase in the density radius can be attributed to a more spread out central concentration of stars due to reduced star formation. The simulation with $v_w = 5000$ km s$^{-1}$ presented an increment of $\sim 10\%$ in stellar density radius. As in the case with $M_{\textit{BH,seed}} = 10^3$ M$_{\odot}$ and $v_w = 1000$ km s$^{-1}$, it produced a negligible difference in the final stellar mass in Fig.~\ref{fig:delta-star} and an accelerated gas depletion (more pronounced here) in Fig.~\ref{fig:mass-tidal}. Therefore, the variation in density radius here can also be related to the presence of an IMBH in the galaxy, with a mass of $10^4$ M$_{\odot}$ in this case. 

These observed trends arise through an indirect influence of AGN feedback on the stellar distribution, as the black hole accretion and feedback for $M_{\textit{BH,seed}} = 10^3, 10^4$ M$_{\odot}$ are not strong enough to critically quench star formation or drive substantial black hole growth (see Fig.~\ref{fig:BH-growth}). However, AGN feedback do alter the star formation loci by modifying gas dynamics, as evidenced by the accelerated gas depletion shown in Fig.~\ref{fig:mass-tidal}.


\subsection{Feedbacks and outflows}

\begin{figure*}
	\includegraphics[width=\textwidth]{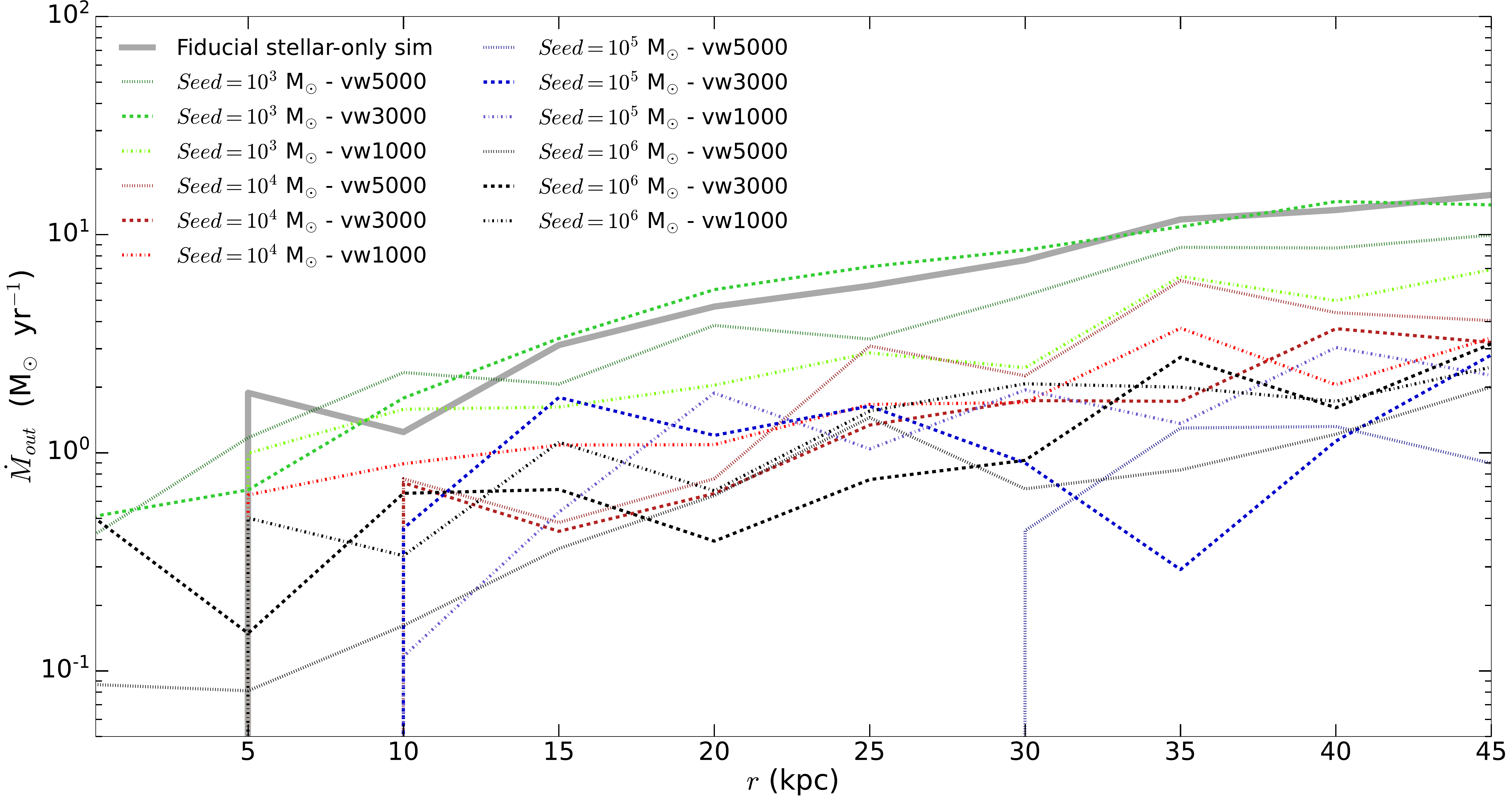}
    \vspace*{0mm}
    \caption{Radial profiles of mass outflow rates estimated for selected simulations with $\epsilon_f = 0.01$.}
    \label{fig:out_rate}
\end{figure*}

Figure \ref{fig:out_rate} displays the mass outflow rates for gas at 13.7 Gyr across the galactic radius for selected simulations with $\epsilon_f = 0.01$. A gas particle was classified as an outflowing particle when its velocity exceeded the escape velocity of the galaxy at the virial radius ($v_r > v_{\text{esc}} \sim24$ km s$^{-1}$) and was directed outward. The outflow rates were then calculated following \citet{Barai2018}

\begin{equation}
    \label{eqn:outflow-rate}
    \dot M_{out}(r) = \sum_{v_{r,i} > v_{esc}} \frac{m_i |v_{r,i}|}{\Delta r} 
\end{equation}

All the curves representing gas mass outflow rates in simulations with AGNs are consistently positioned below the curve for the fiducial simulation with stellar-only feedback, except for the curves corresponding to $M_{\textit{BH,seed}} = 10^3$ M$_{\odot}$ during certain periods. Notably, the curve for $v_w = 3000$ km s$^{-1}$ remained slightly above the fiducial curve for an extended duration. Note that this this curve represents the case of the strongest positive AGN feedback, as depicted in Fig.~\ref{fig:delta-star}, which likely contributes to the observed increase in the outflow rate.

\begin{figure*}
	\includegraphics[width=\textwidth]{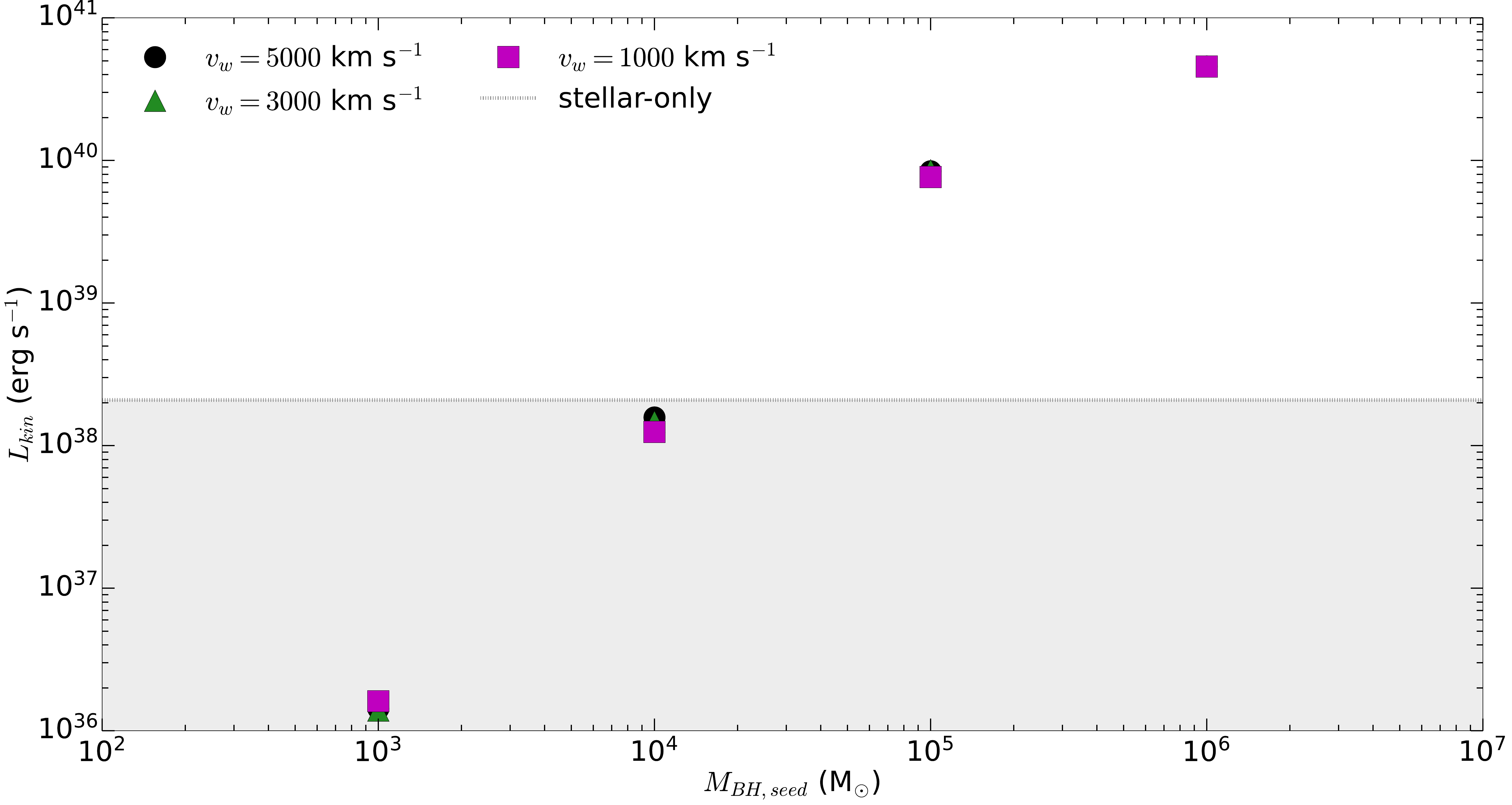}
    \vspace*{0mm}
    \caption{Maximum injected kinetic luminosity estimated for selected simulations with $\epsilon_f = 0.01$.}
    \label{fig:lum_kin}
\end{figure*}

All other curves generally remained below the fiducial case, indicating reduced outflow rates for $M_{\textit{BH,seed}} = 10^3-10^6$ M$_{\odot}$ and all tested velocities. These trends can be attributed to the dominance of stellar feedback as the primary mechanism influencing gas outflow in the simulated galaxies (see also Fig.~\ref{fig:lum_kin}), as cases with more substantial reductions in final stellar mass generally exhibited lower outflow rates. Note, however, that there is not a clear relationship observed among the curves with respect to BH seed mass and AGN wind velocities. This lack of correlation may stem from the significant differences in the galaxies produced in each case, including extreme scenarios such as dark galaxies for $M_{\textit{BH,seed}} = 10^6$ M$_{\odot}$.

\begin{figure*}
	\includegraphics[width=\textwidth]{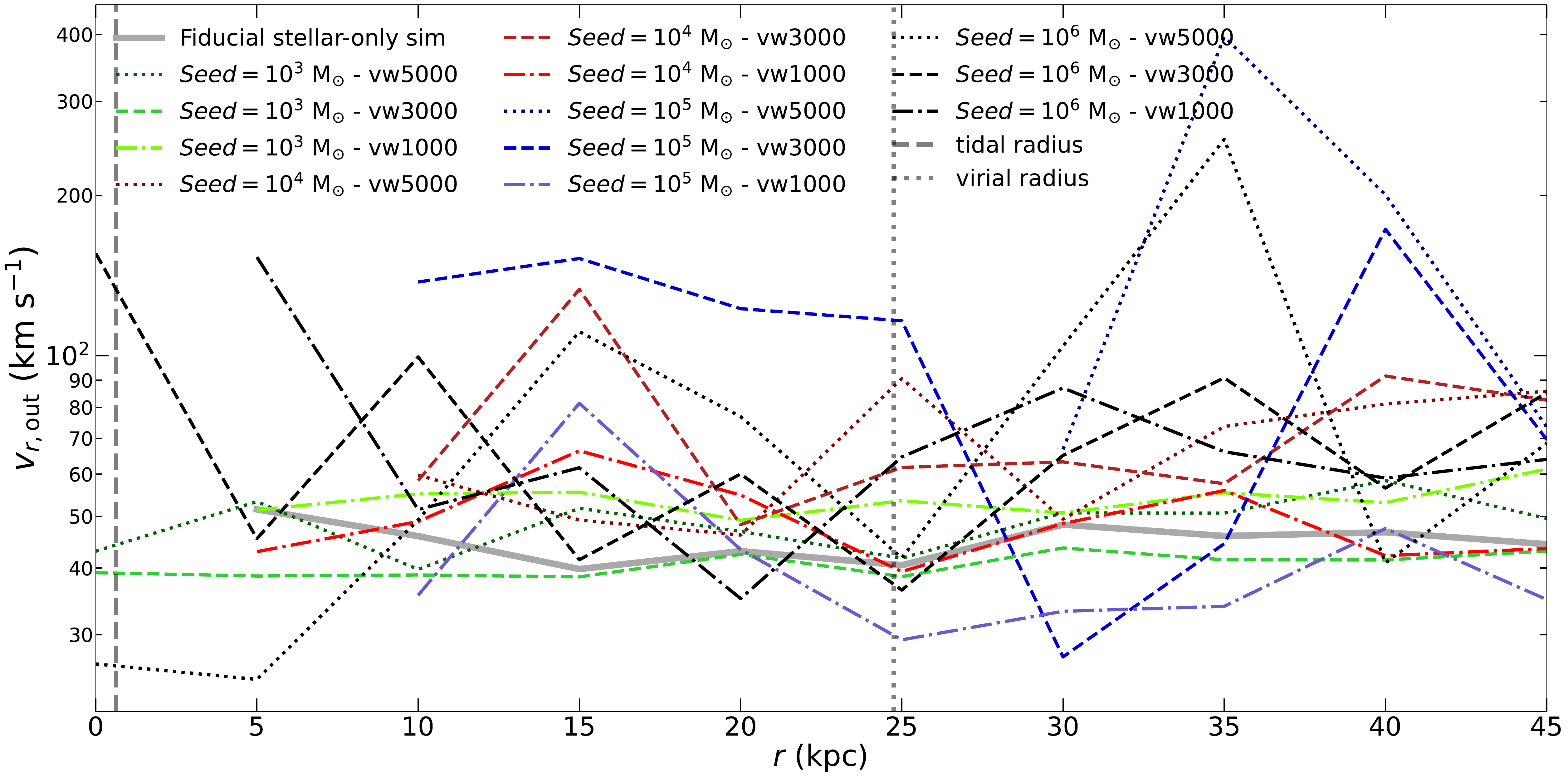}
    \vspace*{0mm}
    \caption{Radial profiles of gas outflow velocities estimated for selected simulations with $\epsilon_f = 0.01$ at 13.7 Gyr.}
    \label{fig:out_vel}
\end{figure*}

The radial profiles of the outflow velocities at 13.7 Gyr are shown in Fig.~\ref{fig:out_vel}, with values ranging between $\sim 20-400$ km s$^{-1}$ for the selected cases. In general, the velocity profiles exhibit higher values than those found in the fiducial stellar-only simulation, at least within a limited galactic radius interval. However, an exception is observed again in the simulation for $M_{\textit{BH,seed}} = 10^3$ M$_{\odot}$ and $v_w = 3000$ km s$^{-1}$, which represents a case of positive feedback as illustrated in Fig.~\ref{fig:delta-star}. This particular simulation exhibits slightly lower outflow velocities across the entire range of galactic radii. This trend suggests that the AGN feedback has reduced the influence of stellar feedback during the galactic evolution. 

One hypothesis considered is that the discrete reduction in outflow velocities for this case could be attributed to the wandering pathways of the black holes (see Fig.~\ref{fig:gas_maps}). These displacements within the galaxy could generate isotropic AGN outflows capable of attenuating gas outflows originating from stellar feedback from the central region, where star formation is concentrated within approximately $500$ pc. This discrete attenuation might then be correlated with the observed increase in star formation in this simulation.

The profiles with the highest velocities were observed for BH seed masses of $M_{\textit{BH,seed}} = 10^5$ and $10^6$ M$_{\odot}$, reaching values of up to $\sim 200-400$ km s$^{-1}$ after the galactic virial radius. Some preliminary reference from more massive dwarfs for comparison with our simulated galaxies may be found in \citet{Manzano2019}, who found outflow velocities in the range $375 - 1090$ km s$^{-1}$ in dwarfs with stellar masses ranging between $4 \times 10^8 - 9 \times 10^9$ M$_{\odot}$ both with and without AGN. More recently, \citet{Liu2024} reported outflow velocities around $230$ km s$^{-1}$ in dwarf galaxies with stellar masses between $10^{8.8-9.4}$ M$_{\odot}$, where AGN is likely the dominant energy source driving these outflows.

To assess the relative impact of the two feedback mechanisms on galactic evolution, we plotted the maximum kinetic luminosity for both stellar and AGN feedbacks in Fig.~\ref{fig:lum_kin}. The maximum values were determined based on the peak star formation rate \citep{Hazenfratz2024} and black hole accretion rates (Eq.~\ref{eq:outflow-rate}).

In Fig.\ref{fig:lum_kin}, the maximum kinetic luminosity for stellar feedback is approximately $2 \times 10^{38}$ erg s$^{-1}$. Simulations that surpassed this power level at certain points were those with $M_{BH,seed} = 10^5$ and $10^6$ M$_{\odot}$ by more than 1 and 2 orders of magnitude, respectively. Such a substantial increase in power could explain the pronounced suppression of star formation observed in Fig.~\ref{fig:delta-star} for these seeds. For $M_{BH,seed} = 10^4$ M$_{\odot}$, the power values are comparable to those for star formation, already sufficient to induce star formation suppression of up to approximately $70\%$ (for $v_w = 1000$ km s$^{-1}$) as shown in Fig.~\ref{fig:delta-star}. Lastly, for $M_{BH,seed} = 10^3$ M$_{\odot}$, the maximum AGN injected power decreased by approximately 2 orders of magnitude in the simulations, thus justifying the more discrete effects observed on star formation suppression for these black hole seeds.

\subsection{Kinetic vs. thermal AGN feedback}

\begin{figure*}
	\includegraphics[width=\textwidth]{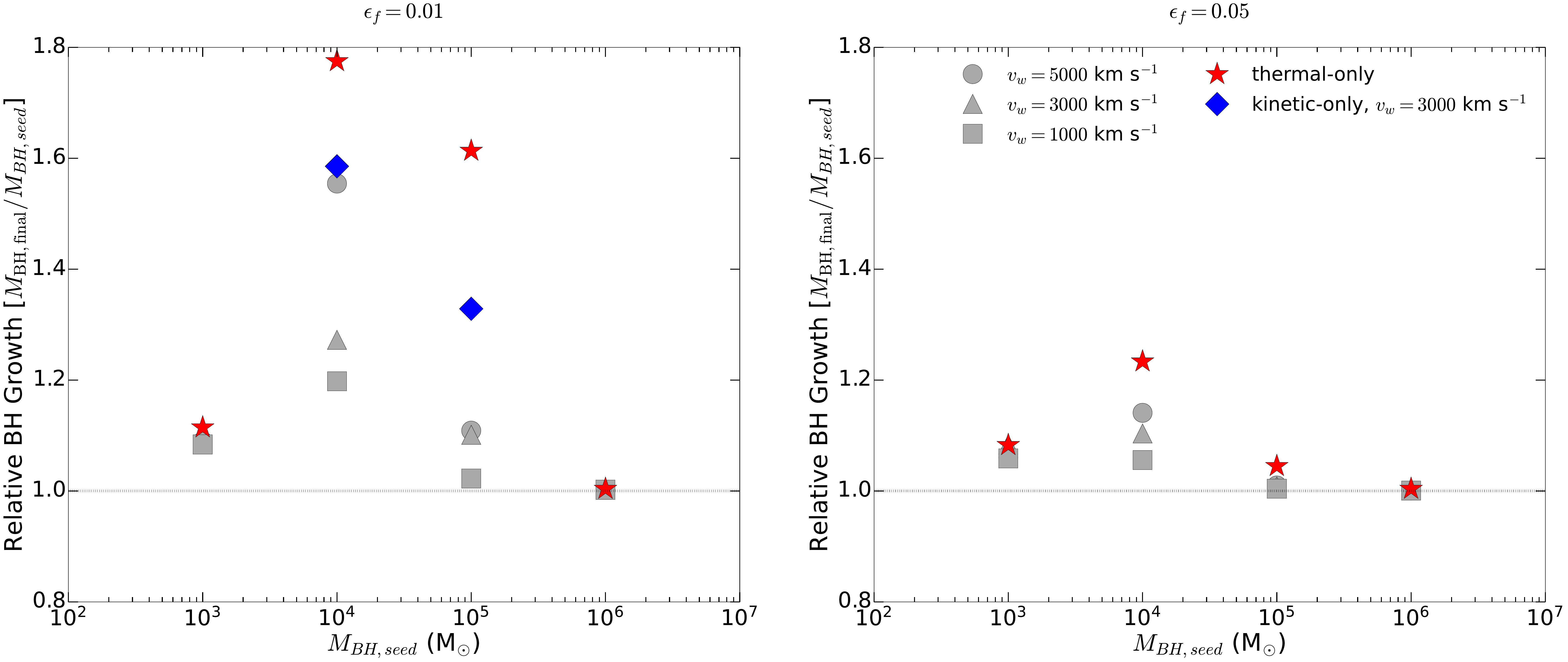}
    \vspace*{0mm}
    \caption{Influence of AGN feedback modes on the growth of black hole seeds in an isolated dwarf spheroidal galaxy.}
    \label{fig:BHgrowth-thermal}
\end{figure*}

To evaluate the impact of various modes of AGN feedback on the overall properties of our simulated galaxy, we investigated simulations that incorporated thermal-only, kinetic-only, and combined thermal and kinetic feedback. Figure \ref{fig:BHgrowth-thermal} illustrates the relative growth of black holes ($M_{\textit{BH,final}}/M_{\textit{seed}}$) for each mode of AGN feedback.

In all scenarios, the inclusion of AGN kinetic feedback reduced the IMBHs growth from black hole seeds when compared to cases with thermal-only feedback. This reduction is particularly noticeable for $M_{\textit{BH,seed}} = 10^4$ and $10^5$ M$_{\odot}$ across both tested feedback efficiencies. The most significant black hole growth is observed for $M_{\textit{seed}} = 10^4$ M$_{\odot}$, yet this enhancement is not substantial enough to alter the final mass magnitude. The ratio of black hole growth remains below 2 in all cases. The increased growth observed for $M_{\textit{seed}} = 10^4$ M$_{\odot}$ can be attributed to an optimal balance between feedback power (forcing gas outwards) and gravitational forces. Notably, for $M_{\textit{seed}} = 10^6$ M$_{\odot}$, the exclusion of the kinetic mode of AGN feedback did not impact the already negligible growth observed.

Two simulations (K4A100E1V3R, K5A100E1V3R) were conducted to examine the impact of removing thermal feedback, with $v_w = 3000$ km s$^{-1}$. These BH seed masses were chosen based on evidence from the literature suggesting the presence of intermediate-mass black holes in this mass range within classical dwarfs similar to Leo II \citep[e.g.,][]{Lora2009, Nucita2013, Manni2015}, with $\epsilon_f = 0.01$ chosen as the feedback efficiency to be further investigated, as discussed in section~\ref{SFH}. The relative growth of BHs for kinetic-only feedback presented intermediate values between the ratios obtained for thermal+kinetic and thermal-only feedback cases. This observation suggests an enhanced effect achieved through the combination of both AGN feedback mechanisms in reducing BH growth via gas accretion in the isolated dwarf galaxy.

\begin{figure*}[h!]
	\includegraphics[width=\textwidth]{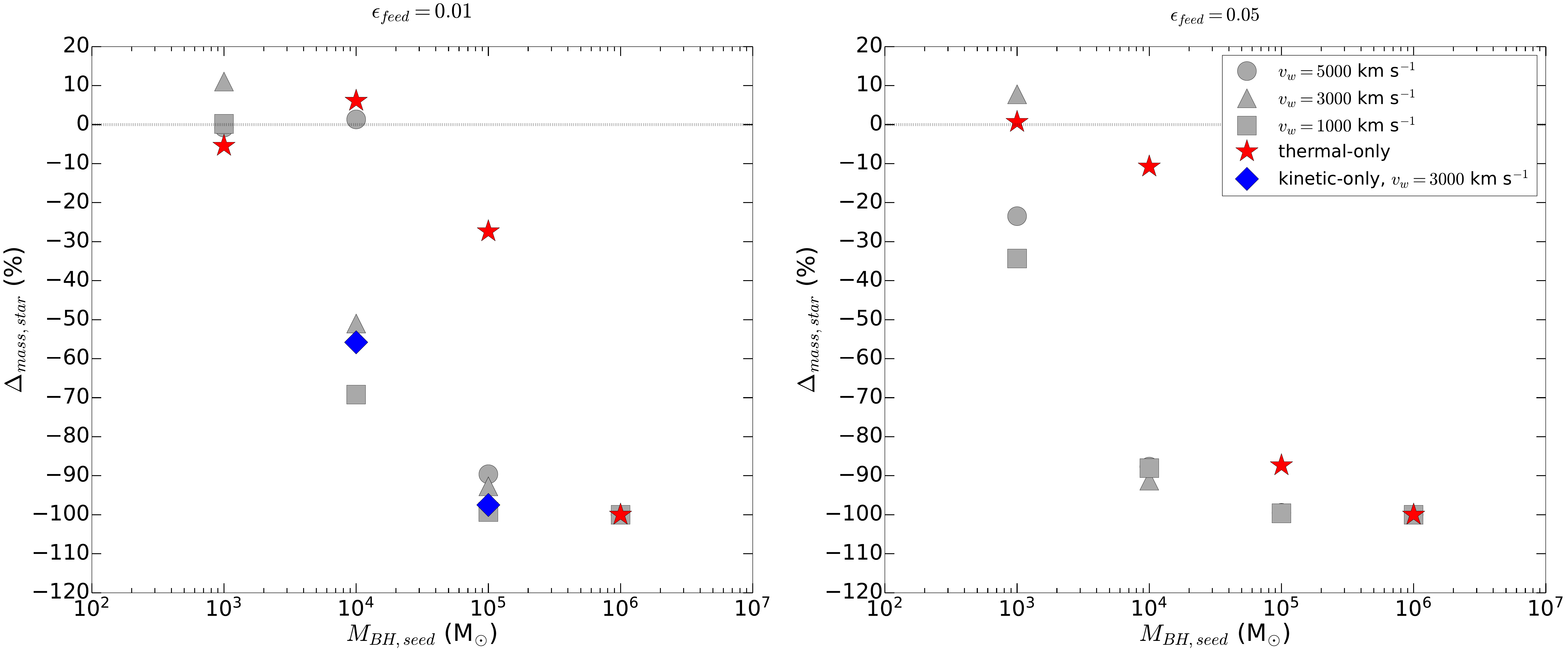}
    \vspace*{2mm}
    \caption{Influence of AGN feedback modes on the final stellar mass of an isolated dwarf spheroidal galaxy.}
    \label{fig:delta-star-thermal}
\end{figure*}

The impact of AGN feedback type and BH seed mass on the final stellar mass formed during the simulations was also examined, as depicted in Figure \ref{fig:delta-star-thermal}.

For BH seeds of $10^4$ and $10^5$ M$_{\odot}$, simulations employing thermal-only AGN feedback exhibited a weaker suppression in the final stellar mass for both efficiencies, with a case of positive AGN feedback observed for $10^4$ M$_{\odot}$. Conversely, for BH seeds of $10^3$ M$_{\odot}$, simulations with thermal-only feedback demonstrated stronger suppression of stellar activity for $\epsilon_f = 0.01$, while showing marginal differences compared to the fiducial stellar-only simulation for $\epsilon_f = 0.05$. This trend suggests that for this mass category, potential positive AGN feedback would primarily be associated with the inclusion of the kinetic feedback mode to generate an adequate degree of gas compression for enhancing star formation. Results for BH seeds of $10^6$ M$_{\odot}$ showed no differences, with complete suppression of star formation in the galaxy.

For simulations incorporating kinetic-only AGN feedback with BH seed masses of $10^4$ and $10^5$ M$_{\odot}$ and $v_w = 3000$ km s$^{-1}$, the exclusion of thermal feedback led to a slightly lower decrease in stellar mass than observed in simulations with combined thermal and kinetic AGN feedback at the same wind velocity. Compared to thermal-only cases, however, the suppression effect was significantly higher, indicating that kinetic feedback is the primary mechanism driving star formation suppression throughout galactic evolution.

\subsection{Influence of the IMBH dynamics on AGN feedback}

In a study with more massive dwarf galaxies, \citet{Bellovary2019} observed that approximately 50$\%$ of the black holes were not centrally located, but instead were wandering within a few kiloparsecs from the galaxy center. In other work, \citet{Ricarte2021} predicted that the most significant BH offsets are likely to originate from SMBHs residing in satellite galaxies. Using the NEWHORIZON cosmological simulations, \citet{Beckmann2023} also found that low-mass black holes ($\sim 10^{4-5}$ M$_{\odot}$) in low-mass galaxies struggled to grow and to remain near the center of low-mass galaxies. From an observational perspective, candidates for wandering black holes have been identified in more massive dwarf galaxies than those considered in this study, confirming the predictions of simulations \citep[e.g.][]{Mezcua2020, Reines2020, Sargent2022}.

These BH displacements were also identified in our simulations, despite the absence of mergers due to our isolated framework. To investigate potential influences of this wandering trend on the AGN feedback effects and, consequently, on the evolution of our benchmark dwarf spheroidal galaxy, we incorporated modifications into some simulations.

\begin{figure*}[h!]
    \includegraphics[width=\textwidth]{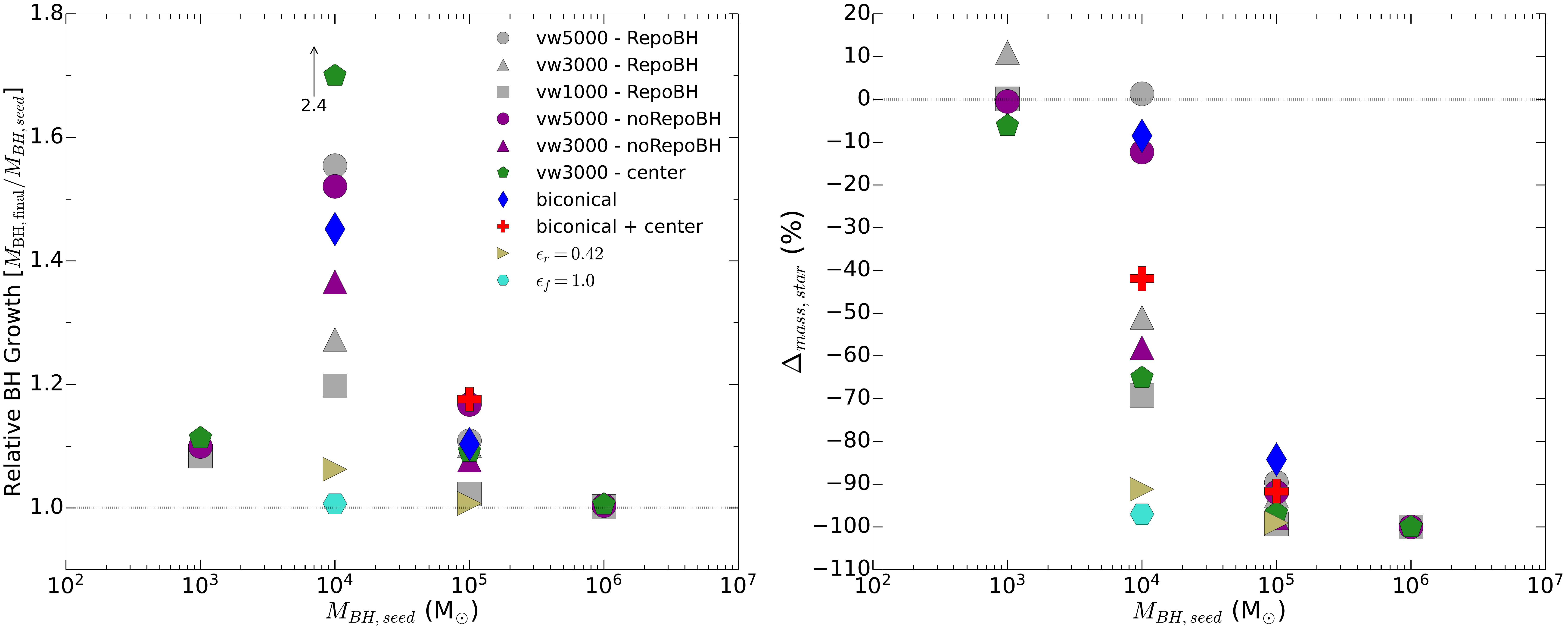}
    \vspace*{2mm}
    \caption{Impact of modifications in black hole dynamics, AGN wind geometry, and radiative and feedback efficiencies on black hole growth and the final stellar mass of an isolated dwarf spheroidal galaxy.}
    \label{fig:BHgrowth-combo}
\end{figure*}

Firstly, we conducted tests comparing the standard method of repositioning black holes at the minimum of the gravitational potential, which was employed in all simulations explored up to this point, with simulations where this option was disabled. Additionally, we tested an extreme scenario where the BH position is frozen at the center of the galaxy to analyze how its dynamics may affect its own growth and influence on the dwarf host galaxy. The results are depicted in Figure \ref{fig:BHgrowth-combo}.

Turning off the repositioning of the black hole at the gravitational potential minimum, thereby allowing it to wander more freely through the galaxy, had little impact on its growth when comparing the same velocities on the left panel of Fig.~\ref{fig:BHgrowth-combo}. However, a relatively stronger impact of removing the black hole repositioning was observed for star formation quenching, as shown on the right panel of the same figure for black hole seeds of $10^4$ M$_{\odot}$. For the two AGN wind velocities tested, the quenching of star formation was more intense with the black hole repositioning turned off, resulting in additional quenching of approximately $-7\%$ and $-14\%$ for $v_w = 3000$ and $5000$ km s$^{-1}$, respectively. For the other black hole seeds, the differences were negligible, not exceeding $5\%$ in any case. This trend can be explained by the reduced influence of black holes with mass of $10^3$ M$_{\odot}$ in the galaxy, which diminishes the importance of black hole dynamics in galactic evolution. Additionally, there is already strong quenching for black holes with masses of $10^5$ and $10^6$ M$_{\odot}$, for which black hole dynamics might contribute less ($\lesssim 3\%$ for $M_{\textit{BH,seed}} = 10^5$ M$_{\odot}$).

When the black hole position is frozen at the defined galaxy center at coordinates (x,y,z) = (0,0,0), more striking differences were observed, but only for $M_{\textit{BH,seed}} = 10^4$ M$_{\odot}$. The black hole's relative growth exceeded that of any case with no freezing and the three AGN wind velocities employed. When comparing the same velocities ($v_w = 3000$ km s$^{-1}$), the growth was approximately $90\%$ higher for the frozen black hole. This increase in black hole growth can be attributed to the null black hole velocity relative to the gas particles, which increases the accretion rate (Eq.~\ref{eq:bondi}), and to higher gas densities that generally occur in the more central regions of the simulated dwarf spheroidal. Regarding the suppression in final stellar mass, it promoted an additional quenching of approximately $14\%$ when comparing the same AGN wind velocity, which can be attributed to the black hole's higher accretion rates and, consequently, higher energy input by AGN feedback into the interstellar medium.

For black hole seed mass of $10^3$ M$_{\odot}$, freezing the BH at the center turned the positive feedback case into suppression of around 8$\%$ in the final stellar mass. For black hole seed masses of $10^5$ and $10^6$ M$_{\odot}$, the impact of freezing the black hole dynamics was negligible, due to the already critical star formation quenching observed under any dynamical scenario.

Clearly, freezing the black hole dynamics represents an artificial scenario in the black hole and galactic evolution. However, it can serve as an extreme case opposite to a scenario where the black hole wanders more freely, albeit without an explicit formulation to treat the dynamical friction of these compact objects in dwarf galaxies. We may infer that once this treatment is considered, the influence of the intermediate-mass black hole on the evolution of our dwarf spheroidal galaxy could exhibit intermediate values for black hole growth and stellar mass suppression between these extreme cases. Note, however, that due to our intermediate resolution, our simulations can indirectly capture some effects of dynamical friction at scales larger than the gravitational softening length (typically $\gtrsim 70$ pc). Therefore, the trajectories of the IMBHs, spanning up to two orders of magnitude greater than the gravitational softening, are not considered artifacts in our simulations, even without an explicit formulation of dynamical friction.

\begin{figure*}
    \includegraphics[width=\textwidth]{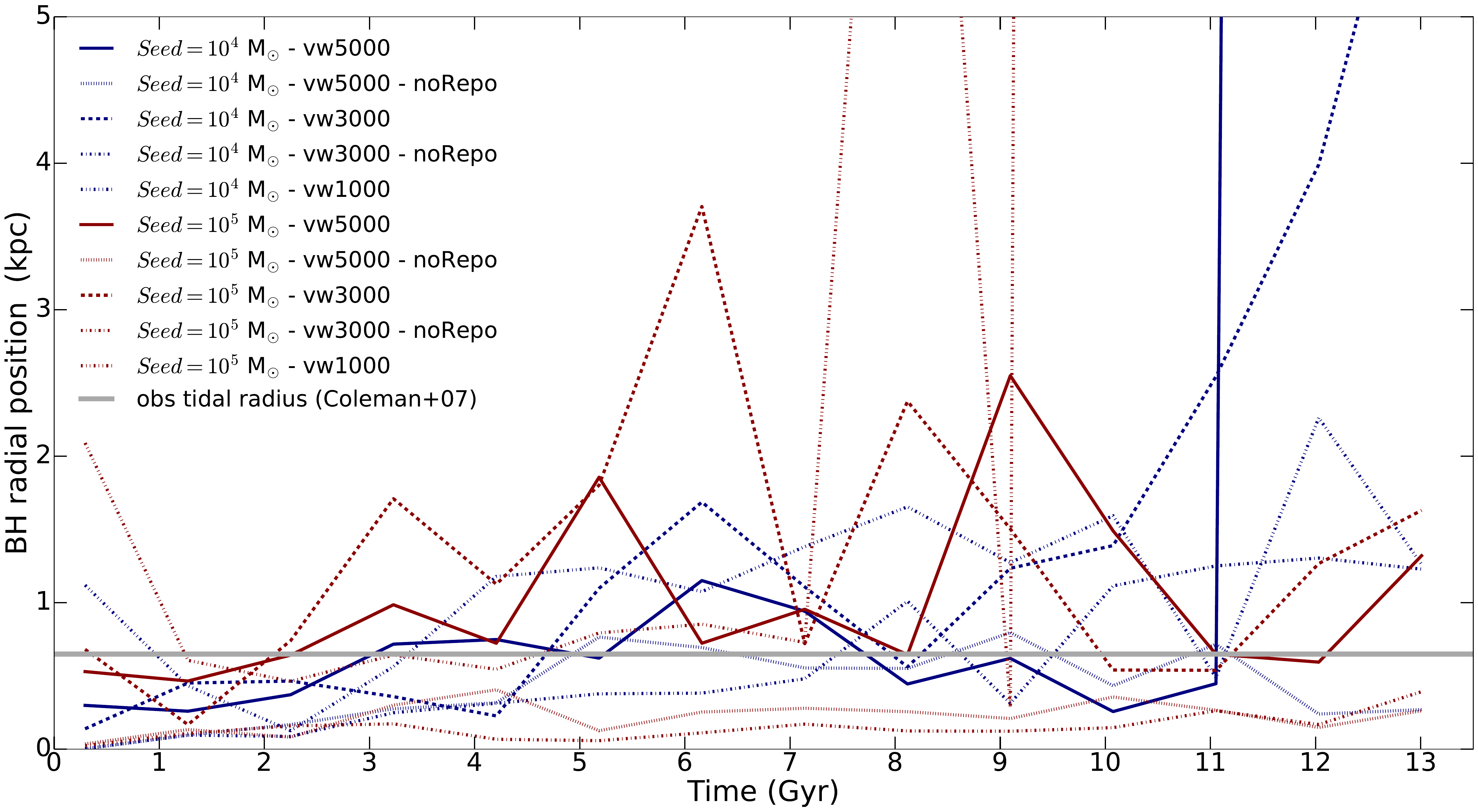}
    \vspace*{2mm}
    \caption{Black hole displacements over time for selected simulations.}
    \label{fig:bh-position}
\end{figure*}

To illustrate the displacements of the black hole throughout the galaxy, its position over time is depicted in Fig.~\ref{fig:bh-position}. A reference value for the gravitational softening length, assuming $1/50$ of the mean particle separation, would be 70 pc. However, for the simulation of our low-mass galaxy, we employed the adaptive gravitational softening length method proposed by \citet{Iannuzzi2011}, which may further reduce this value.

For most simulations, the displacements of the black holes are limited to approximately $2$ kpc. This range is two orders of magnitude higher than the gravitational softening lengths, indicating that numerical artifacts are unlikely to dominate the wandering trajectories of black holes. However, two simulations stand out: one involving a black hole mass of $10^4$ M$_{\odot}$ and $v_w = 5000$ km s$^{-1}$, and another featuring a black hole mass of $10^5$ M$_{\odot}$ and $v_w = 1000$ km s$^{-1}$. In both scenarios, despite employing a repositioning scheme, the black holes were expelled from their host galaxies after 9 Gyr and did not return to the galactic center. By the end of the simulations, the black holes were situated at radial distances exceeding 150 kpc. Notably, despite being beyond the central 5 kpc region, the black hole in the simulation with a mass of $10^4$ M$_{\odot}$ and $v_w = 3000$ km s$^{-1}$ remained within the galactic virial radius. 


The almost complete depletion of stellar formation for the $10^5$ M$_{\odot}$ seed could, in principle, be invoked to justify the higher probability of this event. This could be attributed to the diminished gravitational potential in the central regions of the galaxy, allowing the black hole to wander more freely. However, this argument may not hold for the simulation with a $10^4$ M$_{\odot}$ seed, as it showed almost negligible influence on the final stellar mass formed in the galaxy (see Fig.~\ref{fig:delta-star}). This observation suggests that intermediate-mass black holes could be ejected from their host galaxy even in the absence of a merger event or tidal interactions with neighboring halos. However, further tests should be conducted at higher resolution and with a detailed treatment of dynamical friction to draw more conclusive results.

Disabling the BH repositioning scheme did not show a clear impact on the displacements of black holes for seeds of $10^4$ M$_{\odot}$, but it consistently reduced the wandering radius for seeds of $10^5$ M$_{\odot}$. However, due to the significant suppression of stellar formation in these simulations, further testing should be conducted with these black holes in more massive dwarfs, where the suppression of stellar activity would be diminished, in order to isolate the effect of the repositioning scheme.

\begin{figure*}
    \includegraphics[width=\textwidth]{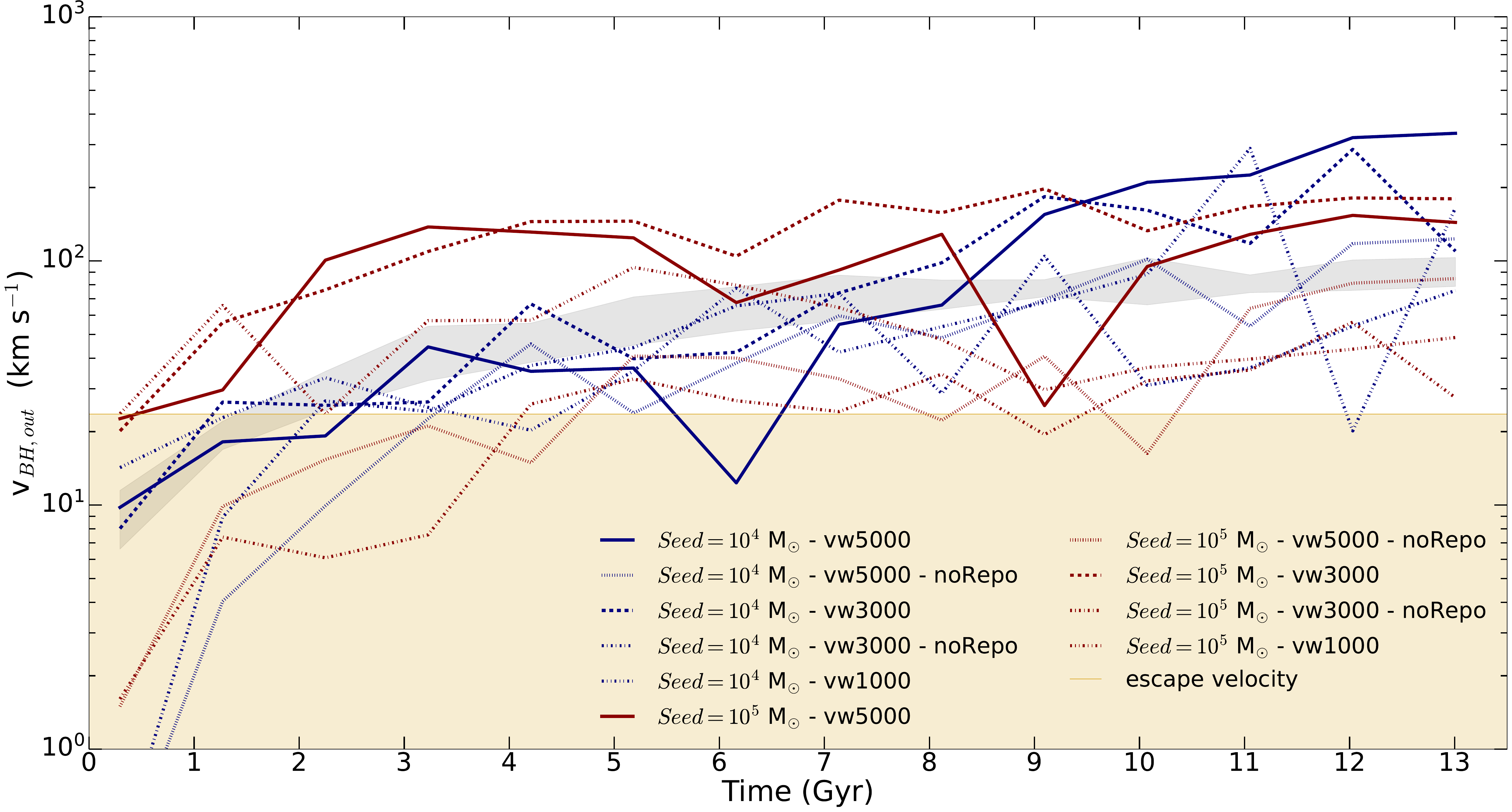}
    \vspace*{2mm}
    \caption{Time evolution of radial velocities for simulated black holes in selected simulations. The gray-shaded area represents the range of median radial velocities observed for star particles.}
    \label{fig:bh-radial-vel}
\end{figure*}

As a final kinematic characterization of our simulated black holes, their radial velocities are depicted in Fig.~\ref{fig:bh-radial-vel}. It was observed that after 5 Gyr of galactic evolution, all the BHs in the simulations essentially exhibit velocities higher than the escape velocity of our simulated dSph, being similar or lower than the median radial velocities for star particles.

In one of the simulations where the black hole was expelled from the galaxy, featuring a mass of $10^4$ M$_{\odot}$ and $v_w = 5000$ km s$^{-1}$, the BH exhibited the highest radial velocities, reaching approximately $\sim 330$ km $s^{-1}$. This velocity could potentially be linked to its ejection from the galaxy. However, in another similar simulation with a black hole mass of $10^5$ M$_{\odot}$ and $v_w = 1000$ km s$^{-1}$, the resulting black hole had a velocity of approximately $\sim 50$ km $s^{-1}$ at the end of simulation. While this velocity exceeds the galaxy's escape velocity, it is not exceptionally high when compared to the median velocities of star particles. These trends underscore the necessity of analyzing black hole dynamics with an explicit dynamical friction scheme at higher resolutions to further evaluate the physical plausibility of their ejection and to rule out numerical artifacts. As shown in the convergence test presented in the Appendix, the black hole was also expelled when a simulation with higher resolution was employed.

\begin{figure*}
    \includegraphics[width=\textwidth]{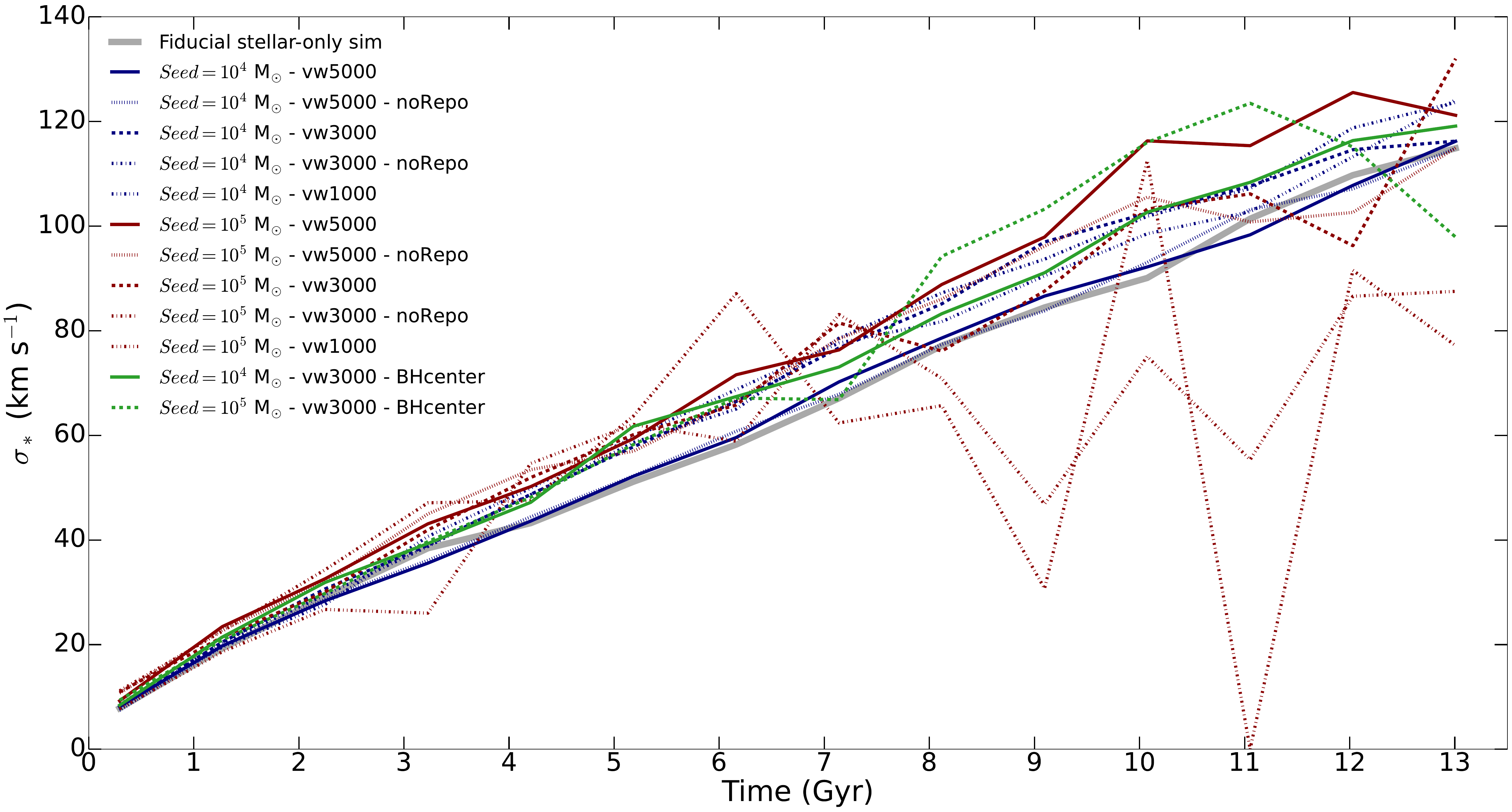}
    \vspace*{2mm}
    \caption{Time evolution of radial velocity dispersion for star particles in selected simulations.}
    \label{fig:vel-dispersion}
\end{figure*}

Finally, to assess whether the presence of an IMBH left a dynamical signature in the stars of the simulated galaxies, we calculated the velocity dispersion for star particles within the central kiloparsec, as plotted in Fig.~\ref{fig:vel-dispersion}. The interest in this parameter stems from the possibility that if the IMBH is not directly observable due to its low accretion activity and luminosity, it might still be detectable based on the imprint from its dynamical interaction with stars.

The overall trend in velocity dispersion showed an increase between approximately $10-120$ km s$^{-1}$ over the course of galactic evolution. Most curves displayed similar behaviors, slightly above that of the fiducial stellar-only simulation. Differences in values could reach up to around 20$\%$. In principle, this suggests a potential dynamical influence stemming from the presence of IMBHs throughout galactic evolution. The highest values of velocity dispersion were observed for seeds of $10^5$ M$_{\odot}$ with $v_w = 5000$ km s$^{-1}$ and $v_w = 3000$ km s$^{-1}$. However, the higher sensitivity of such simulations may be more closely associated with the significant suppression of star formation within the galaxy (over 90$\%$ as shown in Figs.~\ref{fig:delta-star} and \ref{fig:BHgrowth-combo}), resulting in fewer star particles and thus a heightened susceptibility to dynamical interactions from wandering BHs due to the reduced gravitational potential in the central galactic regions.

The exception for this trend are the simulations with black hole seeds of $10^5$ M$_{\odot}$ and $v_w = 3000$ km s$^{-1}$ without BH repositioning and $v_w = 1000$ km/s with BH repositioning. However, these discrepancies and higher oscillations in the velocity dispersion values may be attributed to the low number of star particles for adequate characterization, as these simulations displayed almost complete suppression of stellar formation (Fig.~\ref{fig:delta-star}).

Caution must be maintained when interpreting the velocity dispersion curves depicted in Fig.~\ref{fig:vel-dispersion}. The final values for this feature appear excessively high for dwarf spheroidal galaxies ($\sim$ 10 km s$^{-1}$), more akin to those observed in low-mass elliptical galaxies \citep{Walker2007, Mateo2008, Cimatti2020, Ferre2021}. Simulations at higher resolutions are recommended to more accurately address this discrepancy. The gravitational influence radius, defined as $r_h = GM_{\text{BH}} / \sigma$ \citep{Peebles1972}, was estimated to be less than approximately 2 pc for black hole seeds of $10^3$ M$_{\odot}$ and $10^4$ M$_{\odot}$, thus remaining unresolved in our simulations. However, for heavier seeds, this radius would range from approximately $1-200$ pc, potentially being resolved, as the gravitational softening lengths can be as small as $\lesssim 70$ pc.

A lack of dynamical signatures from the presence of intermediate-mass black holes was noted in cosmological simulations conducted by \citet{Bellovary2021}, where no anomalous dynamical states potentially detectable by an observer were found. In their investigation, the stellar velocity dispersion exhibited temporal oscillations, with values ranging from approximately 10-30 km s$^{-1}$. The absence of influence from IMBHs on this parameter was attributed to the limited number of star particles within the radius of influence of the black holes, which was resolved in their simulations.

\subsection{AGN wind geometry influence}

Bipolar jet-like gas outflows were not identified in simulations with either the thermal-only or the thermal+kinetic AGN feedback. To investigate whether a more angularly concentrated wind injection could induce these features in our simulated dwarf galaxy, we conducted simulations with biconical wind injection within projected angles of $30^\circ$. The results regarding its impact on black hole relative growth and final stellar mass are displayed in Fig.~\ref{fig:BHgrowth-combo}, for black hole seed masses of $10^4$ and $10^5$ M$_{\odot}$ and an AGN wind velocity of $3000$ km s$^{-1}$.

No clear bipolar outflows were observed in the 2D maps of gas properties, as shown in Fig.~\ref{fig:gas_maps}. Furthermore, the implementation of biconical AGN wind injection did not significantly alter black hole growth. However, notable differences were observed in the final stellar mass, with biconical injection reducing the impact on star formation by $\sim 40\%$ for a black hole seed of $10^4$ M$_{\odot}$ and by $\sim 10\%$ for a black hole seed of $10^5$ M$_{\odot}$. These varying trends suggest that the impact of more localized or distributed AGN feedback energy throughout the galaxy may depend on the mass of the black hole, being less significant for heavier black holes.

The absence of visible bipolar outflows in the simulations, whether with isotropic or bipolar AGN wind injection, can be attributed to the low accretion rates observed in this dwarf galaxy. Across all simulations, the final masses of the black holes remained within the same order of magnitude as their initial seeds. These results agree with findings from \citet{Barai2019}, where simulations did not exhibit clear evidence of structured black hole feedback-driven gas outflows. Notably, outflow signatures were only observed around the most massive BHs, with masses exceeding $10^6$ M$_{\odot}$, manifesting as shock-heated, low-density gas.

\subsection{Radiative efficiency influence}

While a fiducial radiative efficiency of $\epsilon_r = 0.1$ is commonly adopted in cosmological simulations, it is important to note that this value represents an average over a wide range of astrophysical conditions and does not correspond to the exact efficiency of an isolated Schwarzschild black hole. The choice of $\epsilon_r = 0.1$ here and related works \citep[e.g.,][]{Barai2018} is often motivated by practical considerations and the need to parameterize complex accretion and emission processes within a simplified subgrid model. Detailed studies of black hole radiative efficiency, including those with spin, indicate that efficiency can vary significantly depending on parameters such as accretion rate, accretion disk geometry, and magnetic field strength 
\citep[e.g.,][]{Shakura1973, Narayan1995, Mckinney2004, Tchekhovskoy2011}.

Although the details of black hole spin evolution in dwarf galaxies are unknown, we conducted a test assuming a Kerr black hole with maximum spin, for which the radiative efficiency would be $\epsilon_r \sim 0.42$ \citep[e.g.,][]{Schnittman2006}. The results for this test with a black hole seed of $10^4$ M$_{\odot}$ are presented in Fig.~\ref{fig:BHgrowth-combo}.

It was observed that the simulation using the maximum value of radiative efficiency practically inhibited black hole growth and led to almost complete suppression of star formation. Since a black hole mass of $10^4$ M$_{\odot}$ is a plausible mass for a black hole at the center of a galaxy similar to Leo II \citep[e.g.,][]{Lora2009, Jardel2012, Nucita2013, Manni2015, Reines2016, Baldassare2017, Mezcua2020, Greene2020}, this extreme value for the radiative efficiency may be deemed unrealistic. Note that intermediate values for this radiative efficiency would necessitate even lower AGN feedback efficiencies ($\epsilon_f < 0.01$) to prevent total suppression of star formation in the host dwarf galaxy.

\subsection{Accretion factor influence} \label{accretion_fac}

In our simulations, the Bondi radius is never resolved, with the spatial resolution being greater by 7 times up to 2 orders of magnitude than this parameter over time. Therefore, we conducted tests to assess the effect of the accretion factor present to compensate for this discrepancy in Eq. \ref{eq:bondi}. A detailed discussion on this topic can be found in \citet{Booth2009}. Table \ref{tab:acc_factor} presents key results regarding the relative black hole growth and depletion in the final stellar mass for different combinations of black hole seed mass and accretion factors.

\begin{table}
	\centering
	\caption{Influence of the accretion factor on black hole growth and stellar formation}
	\label{tab:acc_factor}
        \begin{tabular}{cccccc} 
		\hline
	    Simulation & $M_{\textit{BH,seed}}$ (M$_{\odot}$) & $\alpha$ & $v_{\textit{wind,AGN}}$ (km s$^{-1}$) & $M_{\textit{BH,final}}/M_{\textit{BH,seed}}$ & $\Delta M_{\textit{stellar}} (\%)$\\
		\hline
		TK4A100E1V3R & $10^4$ & 100 & 3000 & 1.27 & -51\\
        TK4A10E1V3R  & $10^4$ & 10  & 3000 & 1.06 & -1\\
        TK4A1E1V3R   & $10^4$ & 1   & 3000 & 1.01 & -31\\
        TK5A100E1V3R & $10^5$ & 100 & 3000 & 1.10 & -93\\
        TK5A10E1V3R  & $10^5$ & 10  & 3000 & 1.10 & -89\\
        TK5A1E1V3R   & $10^5$ & 1   & 3000 & 1.02 & -66\\
        TK6A100E1V5R & $10^6$ & 100 & 5000 & 1.00 & -100\\
        TK6A1E1V5R   & $10^6$ & 1   & 5000 & 1.03 & -90\\         
		\hline
		\end{tabular}
\end{table}

Table~\ref{tab:acc_factor} shows that reducing the accretion factor results in a decrease in relative black hole growth for seeds of $10^4$ M$_{\odot}$, but has minimal influence for heavier seeds. Concerning the variation in final stellar mass, reducing the accretion factor generally leads to a reduction in stellar activity quenching due to AGN feedback. However, despite the decrease in influence of the dwarf AGN with a reduction in the accretion factor, as expected, the overall conclusions remain valid, with higher depletion of stellar activity observed for black hole seeds of $10^5$-$10^6$ M$_{\odot}$.

\subsection{Putative IMBH in Leo II}

This section is driven by the following inquiry: what would be the upper limit on the black hole mass that could exist in Leo II, such that its influence on galactic evolution might remain obscured in simulations or even in observational assessments of its star formation history and associated characteristics?

Based on the results indicating star formation suppression and more tenuous related effects, it is plausible to infer that an IMBH up to $10^4$ M$_{\odot}$ could exist in a dwarf spheroidal galaxy with total mass of approximately $10^9$ M$_{\odot}$ without exerting significant observational influence on the characterization of its star formation history. This conjecture is supported by the notion that the impact of an IMBH up to $10^4$ M$_{\odot}$ on the final stellar mass may be limited to around $50\%$ for an AGN feedback efficiency of $\epsilon_f = 0.01$, assuming certain conditions for AGN feedback in Fig.~\ref{fig:delta-star}. 

From an observational standpoint, the detection of any black hole in the range of $10^3$-$10^4$ M$_{\odot}$ in Local Group dwarfs would likely necessitate the characterization of stellar proper motions using extremely large telescopes, as suggested by \citet{Greene2020}, particularly when there is no evidence of AGN activity. Additionally, as of our current knowledge, there is still no evidence for black holes with masses of $10^3$ M$_{\odot}$ in the literature.

\section{Discussion}

In all the scenarios considered in this work, with changes in the feedback implementation, the BH seeds did not grow through gas accretion to more massive black holes in isolated environments. The discrepancies between the results of isolated and cosmological simulations indicate that IMBHs cannot grow substantially in isolated environments, where galactic mergers do not occur to drive gas inflows toward the central galactic regions. Therefore the efficiency of the BH growth in its initial stages (where the Eddington ratio is lower) remains low in isolated galaxies. This is in contrast to the abundant BH growth observed in cosmological simulations comprising different BH seed masses to study IMBHs and SMBHs \citep[e.g.,][]{Volonteri2010, Barai2018, Barai2019}. 

The observed trends may indicate that a more efficient BH seed growth could occur in galaxies which interact more intensely and frequently with larger hosts or even smaller systems where this galaxy could be considered the host for others, such as ultra-faint dwarf galaxies. In our simulations, increasing the AGN feedback efficiency would further suppress BH growth, while reducing this parameter would not guarantee efficient BH growth exceeding 1 dex in mass.


In our fiducial simulation with stellar-only feedback, the estimated stellar feedback efficiency related to the fraction of the initial gas mass reservoir available for star formation is $\sim 1\%$. This low efficiency is expected for low-mass galaxies due to the influence of stellar feedback in theses systems \citep[e.g.][]{Dekel1986, Governato2010, Hopkins2014}. In simulations that include AGN, this efficiency can decrease further for black hole seeds more massive than $10^4$ M$_{\odot}$. An interesting related question is how black holes seeded from lower masses could grow to such or even higher masses without significantly disrupting their host galaxies. In addition to the usual mechanisms, such as direct collapse, interesting possibilities from magnetohydrodynamical simulations were suggested. The fact that the timescales for hyper-Eddington accretion may be shorter than the SFHs of most galaxies, raises the possibility that black holes could grow through these channels (provided the necessary gas densities are achieved) before star formation feedback becomes a significant factor in the evolution of a dwarf galaxy \citep[e.g.][]{Inayoshi2016, Inayoshi2020, Shi2024}.

Regarding the influence of IMBHs on galactic evolution, when we tested whether off-center black holes may have different accretion histories and influence on their host when compared to BHs frozen at galactic center, we found some differences. Particularly, stellar formation quenching was increased, indicating that the wandering nature of BHs in low-mass galaxies could reduce the impact of its feedback.  Note that the wandering trajectories in our simulations, unlike in cosmological simulations \citep[e.g][]{Bellovary2021}, are entirely due to internal dynamical perturbations, and not due to a merger event, since our simulations are isolated. Furthermore, in any dynamical case, there is a noticeable influence from BHs in their hosts for all BH masses tested. It can be a source of positive feedback, leading to a slight increase in star formation for BH masses below $\sim 10^4$ M$_{\odot}$, while generally producing negative feedback thay variably suppresses star formation for higher BH masses.

Although our BH displacements cannot be directly compared with \citet{Bellovary2019,Bellovary2021} ones due to the different simulation conditions, we can observe some differences. In their work, BH displacements predominantly result from minor mergers, whereas in our simulations, dynamics are primarily influenced by the fuzzier and evolving central gravitational potential in dwarf spheroidal galaxies, due to the higher impact of stellar and AGN feedbacks. Despite the generally lower displacements observed here, the impact can be more pronounced given the smaller size and the absence of disks in our simulated dwarf spheroidals. However, we emphasize the need for caution when interpreting the displacement values reported here, due to the relatively intermediate resolution adopted and the lack of a direct treatment of dynamical friction below the resolved scales. A numerical convergence test is provided in the Appendix.

We estimated typical dynamical friction timescales using the following relation \citep{Binney2011}

\begin{equation}
    t_{fric} = \frac{19 \text{~Gyr}}{\ln \Lambda} \left( \frac{r_i}{5 \text{~kpc}}\right)^2 \frac{\sigma}{200 \text{~km s}^{-1}} \frac{10^8 \text{~M}_{\odot}}{M_{BH}}
\end{equation}

where $\Lambda$ was approximated to $\Lambda \approx (M_{gal}/M_{BH})(R_{BH}/R_{gal})$, with $M_{gal}$ and $R_{gal}$ being the mass and radius of the host galaxy, respectively. The parameter $r_i \sim 1$ kpc is the estimated current BH orbital radius. The stellar velocity dispersion was considered between the typical values observationally estimated for LG dwarfs ($\sigma \sim 10$ km s$^{-1}$) and the higher values achieved in our simulations $\sigma \sim 100$ km s$^{-1}$ to establish lower and upper bounds for the dynamical friction timescale.  

We obtained a lower limit for $t_{fric}$ of $\sim 300$ Gyr for $\sim 10^3$ M$_{\odot}$ and $\sim 40$ Gyr for $\sim 10^4$ M$_{\odot}$ BH seeds. This indicates that it is not expected for these BH seeds to sink to the center of the galaxy within the Hubble time, as these seeds do not grow considerably. For $10^5$ M$_{\odot}$ seeds, we estimated $6 \lesssim t_{fric} \lesssim 60$ Gyr, and $1 \lesssim t_{fric} \lesssim 10$ Gyr for $10^6$ M$_{\odot}$ seeds. Despite the possibility, a reduction in the BH orbital radius was not observed for $10^5$ M$_{\odot}$ seeds in our simulations (see Fig.~\ref{fig:bh-position}). The case for the BH seeds of $10^6$ M$_{\odot}$ was not analyzed due to the tendency of total stellar quenching from AGN feedback. 

One of our main research questions is related to whether AGN could influence star formation in dwarf spheroidal galaxies in a similar way compared to the impact it has in the high-mass regime \citep[e.g.][]{Pillepich2018}. In their work with more massive dwarf galaxies, \citet{Barai2019} found that only BH masses higher than $10^5$ M$_{\odot}$ could suppress star formation in dwarf galaxies. In our work, comprised by a reference galaxy with $1-2$ orders of magnitude lower in virial mass, we found that BHs with $10^4$ M$_{\odot}$ are enough to suppress star formation for the feedback parameters selected. Furthermore, the influence could even start with $10^3$ M$_{\odot}$ in the form of AGN positive feedback.

Despite the general interest in assessing whether dwarf AGNs could be efficient in suppressing star formation, cases of positive feedback were already identified in more powerful counterparts \citep[e.g.][]{Shin2019} and even in dwarf galaxies \citep{Schutte2022}. These cases can be attributed to more moderate AGN feedback, which would affect the gas by compressing regions prone to star formation, instead of triggering generalized outflows. In this work, cases of positive AGN feedback were identified for simulation with black seeds of $10^3-10^4$ M$_{\odot}$.

This study shares the limitations of its companion work \citep{Hazenfratz2024}. It does not take into account reionization effects, time-dependent mass variations of dark matter halos (which could potentially amplify the impact of winds at higher redshifts), characterization of cold, warm and hot phases in the ISM (this would necessitate higher resolutions as in \citet{Gutcke2021}), modifications for the Schmidt law \citep{Kennicutt1998star, Kennicutt1998global} for the low-mass regime of dwarf spheroidal galaxies, variation of the wind velocities with redshift/metallicity and magnetic fields. Furthermore, it should be noted that the BH seeds in the simulations of this work are present from $t = 0$, implicitly assuming that the timescale for any potential formation channel of these seeds is negligible.

Regarding putative dwarf AGNs, the choice of free parameters for the thermal and kinetic feedback models are still based on simulations for more massive galaxies or disky dwarfs, which aimed to reproduce constraints such as the $M_{\textit{BH}}-\sigma$ relation and the distribution of BH masses in cosmological volumes. Hence, the analysis is still limited to exploring lower and upper limits within the parameter space of these models. As noted in other studies \citep[e.g.][]{Koudmani2022}, the results presented here regarding parameter selection for subgrid models should not be considered absolute references, as the calibrated effects of stellar and AGN feedback may be degenerate with other physical processes not included in this work, whether environmental factors or unmodeled internal physical mechanisms.

\section{Summary and Conclusions}

In this study, we incorporated the implementation of an AGN feedback model and the exploration of related physical scenarios of central BHs in a dwarf spheroidal galaxy through isolated SPH simulations executed with a modified version of the GADGET-3 code. The main objectives were to investigate the growth and gas accretion activity of black hole seeds within an isolated environment and evaluate the impact of associated feedback on the evolution of our model galaxy. This study explored the influence of AGN feedback on various aspects, including black hole growth, star formation history, interstellar gas evolution and stellar imprints from the presence of an IMBH. 

Throughout our work, we established comparisons with a fiducial simulation involving solely the implementation of star formation and stellar/SN feedback mechanisms. This included processes such as evaporation of cold clouds due to supernova feedback, and the growth of cold clouds through radiative cooling. Additionally, we incorporated kinetic stellar feedback, utilizing the energy-driven prescription and a chemical evolution model of \citet{Tornatore2007}, adopting a Chabrier IMF. Additional sub-resolution physics considered includes radiative cooling and heating from the photoionizing background, stellar evolution, and chemical enrichment for 11 elements.

Our numerical scheme was then combined with a thermal and kinetic AGN feedback model, considering two different feedback efficiencies ($1\%$ and $5\%$), black hole seeds ranging from $10^3-10^6$ M$_{\odot}$, three AGN wind injection velocities (1000, 3000 and 5000 km s$^{-1}$), as well as modifications affecting the dynamics of the black hole (with repositioning at the gravitational potential minimum, no repositioning and freezing the BH position at the galactic center) and injection of the feedback energy.

Our main findings can be summarized as follows: 

1. We evaluated the evolution of BH seeds within an isolated galaxy setting, where their growth would solely depend on gas accretion, without any influence from mergers or environmental factors. The findings revealed that the intermittent regime of gas accretion onto the BH is consistently reproduced, with Eddington ratios typically remaining below $1\%$ for $\epsilon_f = 0.05$, and below $7\%$ for $\epsilon_f = 0.01$. The increased values observed for the lower efficiency are primarily attributed to the greater availability of gas resulting from the reduced gas outflow rates.

2. We observed that IMBHs do not accrete significant amounts of gas, resulting in mass increments lower than twice their initial mass, regardless of their seed mass, AGN feedback efficiency and wind injection velocity. Our findings align with the overall conclusions drawn by \citet{Bellovary2019}, who noted that massive black holes in dwarf galaxies accrete minimal gas, thereby suggesting that their final masses still reflect the mechanisms involved in seed formation in the early universe.

3. The AGN feedback efficiency significantly influences the evolution of the galaxy, with higher values resulting in more pronounced suppression of star formation within the central region. This suppression is attributed to the reduced availability of gas for star formation. Additionally, our study identified cases of positive AGN feedback for simulations featuring black hole seeds ranging from $10^3$ to $10^4$ M$_{\odot}$.

4. We observed a notable impact of the AGN on the star formation history, which depends on both the BH seed mass and AGN wind velocities. Lower velocities typically led to increased suppression of star formation, particularly evident for BH seed masses of $10^4$ M$_{\odot}$ and $\epsilon_f = 0.01$. Furthermore, gas depletion within the tidal radius was observed to accelerate in most simulations with AGN feedback.

5. The stellar formation suppression observed in simulations with $\epsilon_f = 0.05$ was excessively high for a dwarf spheroidal galaxy in the mass range of Leo II, even for IMBHs of $10^4 - 10^5$ M$_{\odot}$, which was already detected in similar systems. It suggests that the AGN feedback efficiency is lower than $5\%$ in dwarf spheroidals.

6. No star formation activity was detected for
$10^6$ M$_{\odot}$ seeds across all combinations of feedback efficiencies and AGN wind injection velocities. This lack of activity suggests that the presence of an IMBH with this mass would be unlikely in dwarf spheroidal galaxies.

7. Suggestive imprints from the presence of an IMBH in the galaxy were identified in the stellar distribution. These imprints could be attributed to an indirect influence, as the tenuous stellar formation suppression for selected cases with $M_{\textit{BH,seed}} = 10^3, 10^4$ M$_{\odot}$ alone could not explain the differences. However, AGN feedback, although insufficient to significantly quench the galaxy, could alter the star formation loci through changes in gas dynamics, with the latter indicated as an acceleration of gas depletion.

8. In most cases, the presence of an IMBH and its associated feedback within the dwarf galaxy reduced the gas outflow rates and increased the gas outflow velocities up to $\sim 1$ dex.

9. The IMBHs were observed to wander throughout the galaxy over the entire cosmic time, with displacements reaching up to 1-2 kpc in general. Moreover, some simulations suggested the potential for their ejection from the host galaxy, even in the absence of a merger event or tidal interactions with neighboring halos. Typical radial velocities fell between the range of the galactic escape velocity and the median velocities of star particles, though exceptions occurred, particularly with black holes reaching velocities of around $\sim 330$ km s$^{-1}$.

10. To investigate the influence of BH wandering on AGN feedback and, consequently, on the evolution of our benchmark dwarf spheroidal galaxy, we introduced modifications by disabling BH repositioning at the gravitational potential minimum and, as an extreme artificial scenario, freezing the BH at the galactic center. These modifications revealed that turning off the reposition had little impact on BH growth, but led to increased suppression of star formation, particularly for BH seeds of $10^4$ M$_{\odot}$. When the BH was frozen at the galactic center, the suppression further increased when compared to the same AGN wind velocity. Taken together, these results suggest that the wandering character of BHs in dwarf galaxies may reduce the impact of their associated feedback. 

11. Some simulations were conducted with biconical wind injection within projected angles of $30^\circ$ and an AGN wind injection velocity of $3000$ km s$^{-1}$. Similar to isotropic wind injection, no clear bipolar outflows were observed. Furthermore, this method did not significantly alter the black hole growth, but reduced the impact of the AGN on star formation by $\sim 40\%$ for a black hole seed of $10^4$ M$_{\odot}$ and by $\sim 10\%$ for a black hole seed of $10^5$ M$_{\odot}$.

12. An IMBH with mass up to $10^4$ M$_{\odot}$ could potentially exist in a relatively isolated dwarf spheroidal galaxy with total mass of $\sim 10^9$ M$_{\odot}$ without exerting significant observational influence on the characterization of its star formation history. This scenario holds true under certain conditions, specifically for an AGN feedback efficiency of $\epsilon_f = 0.01$. 

In all simulations conducted in this study, the growth of BHs was constrained, reaching a maximum increase of 2.4 times their initial mass. Considering the exploration of the free parameters of the AGN feedback model, our results suggest that IMBHs might experience limited growth in an isolated environment, primarily reflecting their original seed masses. This trend implies that isolated dwarf spheroidal galaxies with detected AGN activity could serve as convenient laboratories for constraining BH seeding mechanisms. An alternative scenario for further growth of these IMBHs may involve galaxy mergers, which have already been identified as a more effective pathway for generating more massive black holes in larger galaxies \citep[e.g.][]{Booth2009, Barai2019}.

Despite the limited growth of the IMBHs, their associated AGN feedback was found to be a substantial source of star formation regulation, impacting the baryon cycle of our isolated dwarf spheroidal galaxy with $M_{halo} \sim 10^9$ M$_{\odot}$. The majority of star formation occurred in the central regions, being suppressed by gas heating and mechanical energy injection. Overall, these findings align with the results of \citet{Koudmani2019, Koudmani2022} for more massive dwarf galaxies. This evidence suggests that, in addition to the pivotal role of stellar feedback, low-efficiency AGN feedback can still impact the evolution of dwarf galaxies down to at least $M_{halo} \sim 10^9$ M$_{\odot}$.  

Our tests and calibrations, which have been conducted through isolated simulations focusing on stellar-only feedback and its combination with AGN feedback in dwarf spheroidal galaxies, will be further evaluated in cosmological framework in an upcoming paper, which will conclude this series.


\section{Acknowledgments}

The authors would like to acknowledge the Coordenação de Aperfeiçoamento de Pessoal de Nível Superior - Brasil (CAPES) - Process Number 88887.484382/2020-00 for the financial support provided for this research. The authors also acknowledge the National Laboratory for Scientific Computing (LNCC/MCTI, Brazil) for granting access to HPC resources on the Santos Dumont supercomputer. PB acknowledges support from the PRIN 2022 PNRR grant: project P2022ZLW4T "Next-generation computing and data technologies to probe the cosmic metal content", funded by the Italian Ministry of University and Research (MUR) Missione 4 - Component C2, under the National Recovery and Resilience Plan (PNRR). Finally, we thank the anonymous reviewers for their feedback and constructive comments, which contributed to improving the quality and clarity of this manuscript.


%




\section*{Data Availability}
 
The simulation data are available upon request.

\appendix

\section{Numerical convergence}

We examined how the resolution adopted in the simulations might influence our findings, as in the case of the first paper of this series with the study of stellar feedback. We conducted a comparative analysis of critical aspects of galactic evolution by employing simulations with different numbers of SPH particles. These tests involved simulations at lower resolution (with half the number of particles) and higher resolution (with twice the number of particles). The initial setup and all the other parameters remained the same in all simulations. Table~\ref{tab:resolution} summarizes selected results for the different runs, where m$_{\text{dm}}$ and m$_{\text{gas}}$ are the dark matter particle mass and initial gas particle mass, respectively. Additionally, we conducted tests with ten times more SPH particles. However, this resulted in stellar particle masses below $10^3$ M$_{\odot}$, potentially compromising the IMF representation. In such instances, stochastic variations within stellar populations may significantly influence stellar feedback \citep[e.g.,][]{Smith2021}.

\begin{table*}[h]
	\centering
        \caption{Influence of mass resolution on the simulations.}
        \label{tab:resolution}
        \resizebox{\textwidth}{!}{\begin{tabular}{ccccccccc} 
		\hline
	    Simulation & Gas particles & m$_{\text{dm}}$ (M$_{\odot}$) & m$_{\text{gas}}$ (M$_{\odot}$) & $\eta$ & $v_{\textit{}{\text{wind}}}$ (km s$^{-1}$) & Residual tidal gas mass (M$_{\odot}$) & Stellar mass (M$_{\odot}$) & $[\text{Fe/H}]_{\text{star}}$\\
		\hline
		TK4A100E1V3R & 20000 & 5.3 $\times 10^4$ & 1.6 $\times 10^4$ & 60 & 96 & 0 & 1.0 $\times 10^6$ & -0.93\\
        lower resolution & 10000 & 1.1 $\times 10^5$ & 3.2 $\times 10^4$ & 60 & 96 & 7.2 $\times 10^4$ & 9.0 $\times 10^5$ & -0.85\\
        higher resolution & 40000 & 2.7 $\times 10^4$ & 8.0 $\times 10^3$ & 60 & 96 & 0 & 1.6 $\times 10^6$ & -0.77\\
        higher resolution & 40000 & 2.7 $\times 10^4$ & 8.0 $\times 10^3$ & 45 & 105 & 0 & 1.2 $\times 10^6$ & -0.88\\
        \hline
        \end{tabular}}
\end{table*}

\begin{figure*}
    \includegraphics[width=\textwidth]{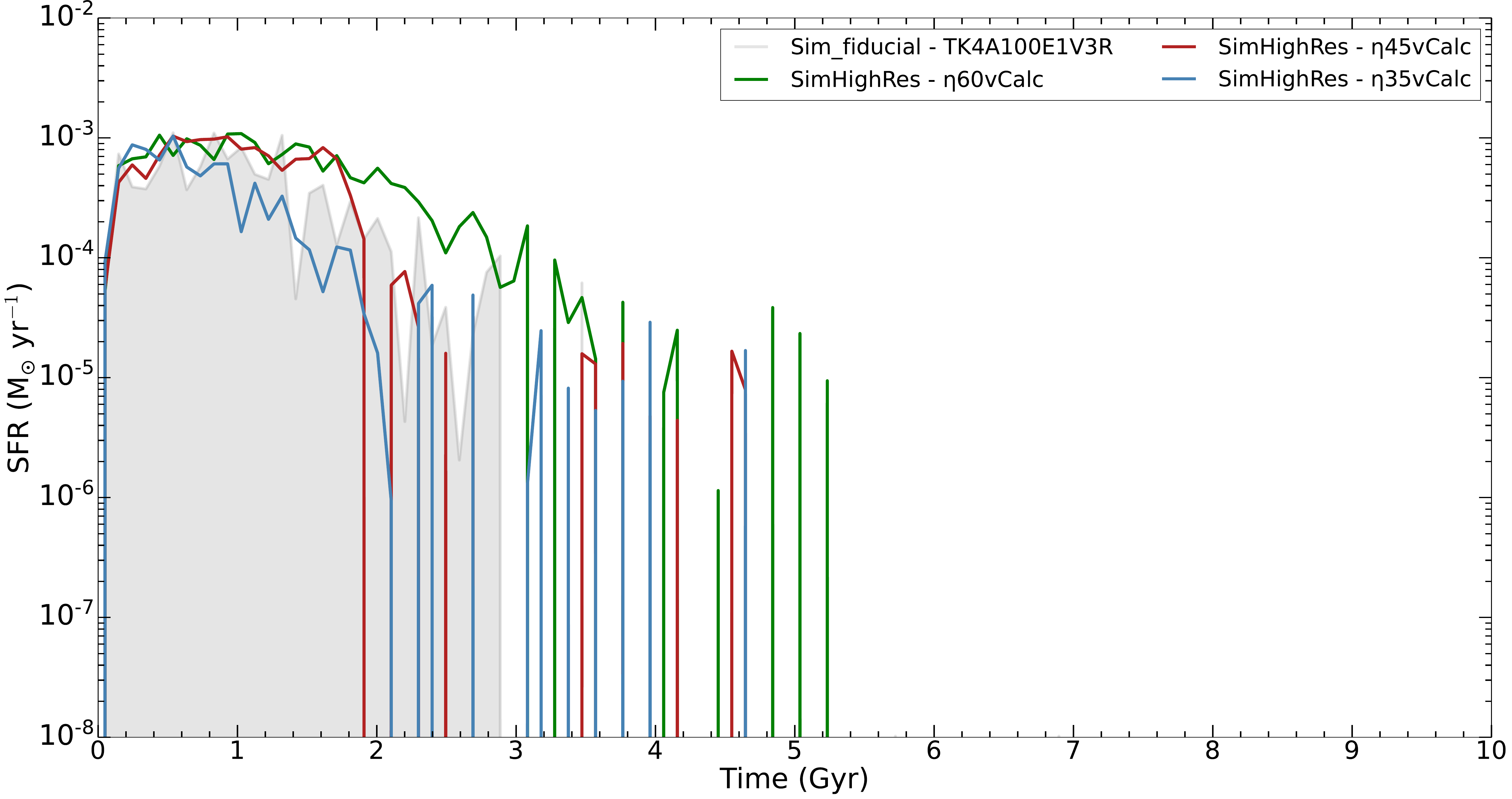}
    \vspace*{-5mm}
    \caption{Star formation history of simulations in the resolution tests.}
    \label{fig:sfr-resolution}
\end{figure*}

Figure~\ref{fig:sfr-resolution} illustrates the star formation histories of the SPH simulations conducted at lower and higher resolutions, contrasted with the fiducial simulation TK4A100E1V3R. In this fiducial simulation, the black hole seed mass is $10^4$ M$_{\odot}$, $\alpha = 100$, feedback efficiency $\epsilon_f = 0.01$, AGN wind injection velocity $v_w = 3000$ km $s^{-1}$, and black hole repositioning.

It was observed that total gas depletion within the tidal radius was only achieved for the simulations at regular and higher resolutions. Regarding the stellar mass formed, the final value exhibited a decrease of approximately $10\%$ for the simulation at lower resolution and an increase of about $60\%$ at higher resolution. Notably, adjusting the value for the mass loading factor in the model for stellar feedback to $\eta = 45$ reduced the variation to around $20\%$. As for the stellar metallicity values, the variation was approximately $-9\%$ for the lower resolution and $-17\%$ for the higher resolution, which decreased to $-5\%$ when $\eta = 45$.

\begin{figure*}
    \centering
    \includegraphics[width=0.9\textwidth]{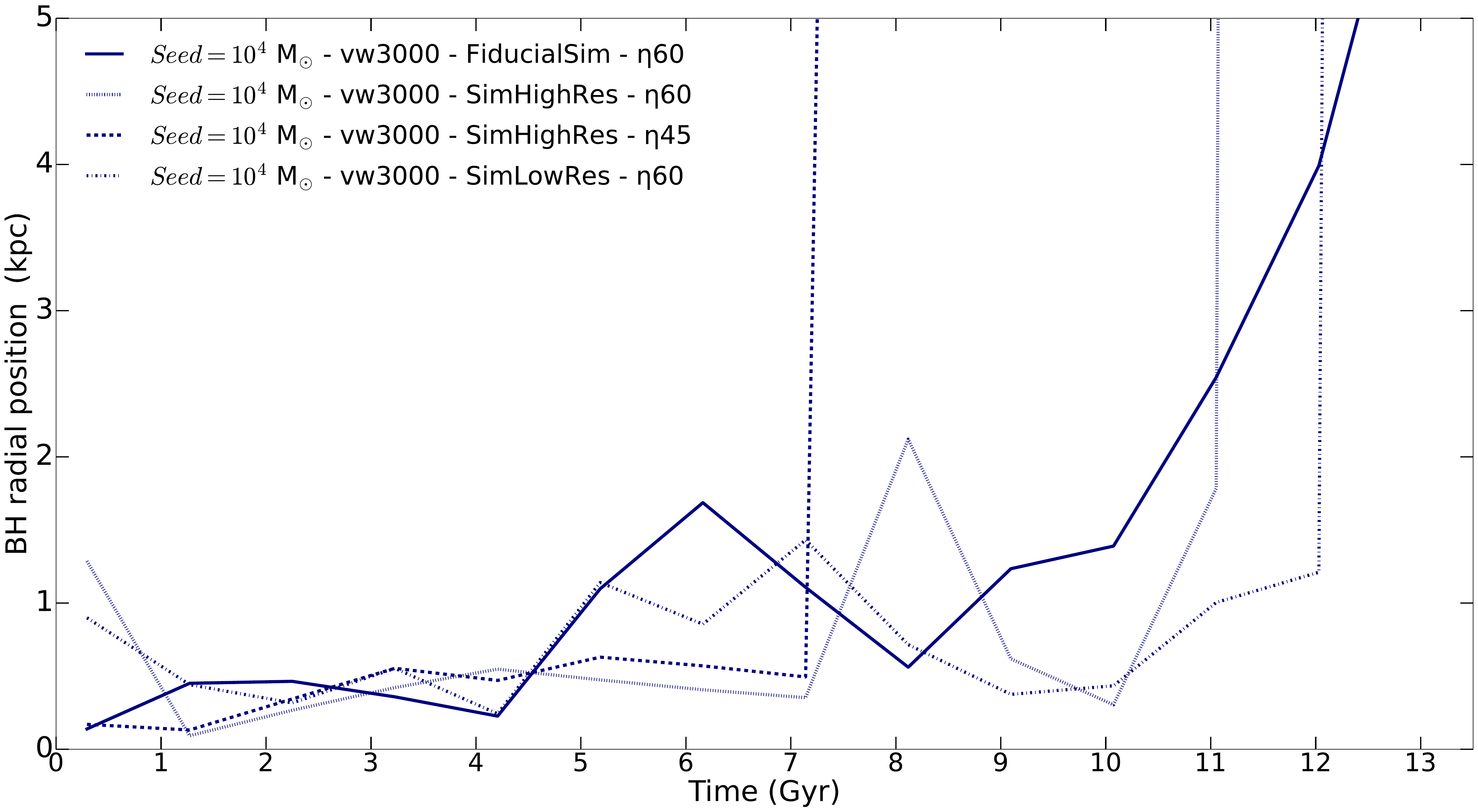}
    \vspace*{-5mm}
    \caption{Black hole displacements over time in the resolution tests.}
    \label{fig:bh-pos-resolution}
\end{figure*}

Figure~\ref{fig:bh-pos-resolution} depicts the displacements of the black holes over time, illustrating the same comparisons with the fiducial simulation TK4A100E1V3R. In this analysis, no differences in the range of values for black hole displacements were observed until the black holes were practically ejected from the galaxy, a phenomenon that occurred in all simulations but at varying cosmic times, occurring earlier for the high-resolution runs.

Similar to the investigation of stellar-only feedback in simulated dwarf spheroidal galaxies, although strict numerical convergence was not attained, consistent results for the star formation history, galactic evolution, and black hole dynamics were obtained across both higher and lower resolutions. Variations in $\eta$ for achieving more similar results were observed to be below 1.5 for the higher resolution case.


\pagebreak
\bibliography{references}{}

\begin{thebibliography}{}
\expandafter\ifx\csname natexlab\endcsname\relax\def\natexlab#1{#1}\fi
\providecommand{\url}[1]{\href{#1}{#1}}
\providecommand{\dodoi}[1]{doi:~\href{http://doi.org/#1}{\nolinkurl{#1}}}
\providecommand{\doeprint}[1]{\href{http://ascl.net/#1}{\nolinkurl{http://ascl.net/#1}}}
\providecommand{\doarXiv}[1]{\href{https://arxiv.org/abs/#1}{\nolinkurl{https://arxiv.org/abs/#1}}}

\bibitem[{Aghanim {et~al.}(2020)Aghanim, Akrami, Ashdown, Aumont, Baccigalupi, Ballardini, Banday, Barreiro, Bartolo, Basak, {et~al.}}]{Planck2018}
Aghanim, N., Akrami, Y., Ashdown, M., {et~al.} 2020, Astronomy \& Astrophysics, 641, A6

\bibitem[{Baldassare {et~al.}(2017)Baldassare, Reines, Gallo, \& Greene}]{Baldassare2017}
Baldassare, V.~F., Reines, A.~E., Gallo, E., \& Greene, J.~E. 2017, The Astrophysical Journal, 850, 196

\bibitem[{Barai \& de~Gouveia Dal~Pino(2019)}]{Barai2019}
Barai, P., \& de~Gouveia Dal~Pino, E.~M. 2019, Monthly Notices of the Royal Astronomical Society, 487, 5549

\bibitem[{Barai {et~al.}(2018)Barai, Gallerani, Pallottini, Ferrara, Marconi, Cicone, Maiolino, \& Carniani}]{Barai2018}
Barai, P., Gallerani, S., Pallottini, A., {et~al.} 2018, Monthly Notices of the Royal Astronomical Society, 473, 4003

\bibitem[{Barai {et~al.}(2014)Barai, Viel, Murante, Gaspari, \& Borgani}]{Barai2014}
Barai, P., Viel, M., Murante, G., Gaspari, M., \& Borgani, S. 2014, Monthly Notices of the Royal Astronomical Society, 437, 1456

\bibitem[{Beckmann {et~al.}(2023)Beckmann, Dubois, Volonteri, Dong-P{\'a}ez, Trebitsch, Devriendt, Kaviraj, Kimm, \& Peirani}]{Beckmann2023}
Beckmann, R., Dubois, Y., Volonteri, M., {et~al.} 2023, Monthly Notices of the Royal Astronomical Society, 523, 5610

\bibitem[{Bellovary {et~al.}(2019)Bellovary, Cleary, Munshi, Tremmel, Christensen, Brooks, \& Quinn}]{Bellovary2019}
Bellovary, J.~M., Cleary, C.~E., Munshi, F., {et~al.} 2019, Monthly Notices of the Royal Astronomical Society, 482, 2913

\bibitem[{Bellovary {et~al.}(2021)Bellovary, Hayoune, Chafla, Vincent, Brooks, Christensen, Munshi, Tremmel, Quinn, Van~Nest, {et~al.}}]{Bellovary2021}
Bellovary, J.~M., Hayoune, S., Chafla, K., {et~al.} 2021, Monthly Notices of the Royal Astronomical Society, 505, 5129

\bibitem[{Binney \& Tremaine(2011)}]{Binney2011}
Binney, J., \& Tremaine, S. 2011, Galactic dynamics, Vol.~13 (Princeton university press)

\bibitem[{Bondi(1952)}]{bondi1952}
Bondi, H. 1952, Monthly Notices of the Royal Astronomical Society, 112, 195

\bibitem[{Bondi \& Hoyle(1944)}]{bondi1944}
Bondi, H., \& Hoyle, F. 1944, Monthly Notices of the Royal Astronomical Society, 104, 273

\bibitem[{Booth \& Schaye(2009)}]{Booth2009}
Booth, C., \& Schaye, J. 2009, Monthly Notices of the Royal Astronomical Society, 398, 53

\bibitem[{Bradford {et~al.}(2018)Bradford, Geha, Greene, Reines, \& Dickey}]{Bradford2018}
Bradford, J.~D., Geha, M.~C., Greene, J.~E., Reines, A.~E., \& Dickey, C.~M. 2018, The Astrophysical Journal, 861, 50

\bibitem[{Bryan {et~al.}(2013)Bryan, Kay, Duffy, Schaye, Vecchia, \& Booth}]{Bryan2013}
Bryan, S., Kay, S., Duffy, A., {et~al.} 2013, Monthly Notices of the Royal Astronomical Society, 429, 3316

\bibitem[{Casertano \& Hut(1985)}]{Casertano1985}
Casertano, S., \& Hut, P. 1985, Astrophysical Journal, Part 1 (ISSN 0004-637X), vol. 298, Nov. 1, 1985, p. 80-94., 298, 80

\bibitem[{Chabrier(2003)}]{Chabrier2003}
Chabrier, G. 2003, Publications of the Astronomical Society of the Pacific, 115, 763

\bibitem[{Cimatti {et~al.}(2020)Cimatti, Fraternali, \& Nipoti}]{Cimatti2020}
Cimatti, A., Fraternali, F., \& Nipoti, C. 2020, Introduction to Galaxy Formation and Evolution (Cambridge,  Cambridge Univ. Press

\bibitem[{Coleman {et~al.}(2007)Coleman, Jordi, Rix, Grebel, \& Koch}]{Coleman2007}
Coleman, M.~G., Jordi, K., Rix, H.-W., Grebel, E.~K., \& Koch, A. 2007, The Astronomical Journal, 134, 1938

\bibitem[{Connor {et~al.}(2023)Connor, Ba{\~n}ados, Cappelluti, \& Foord}]{Connor2023}
Connor, T., Ba{\~n}ados, E., Cappelluti, N., \& Foord, A. 2023, arXiv preprint arXiv:2311.08451

\bibitem[{Correa {et~al.}(2015)Correa, Wyithe, Schaye, \& Duffy}]{Correa2015}
Correa, C.~A., Wyithe, J. S.~B., Schaye, J., \& Duffy, A.~R. 2015, Monthly Notices of the Royal Astronomical Society, 452, 1217

\bibitem[{Dashyan {et~al.}(2018)Dashyan, Silk, Mamon, Dubois, \& Hartwig}]{Dashyan2018}
Dashyan, G., Silk, J., Mamon, G.~A., Dubois, Y., \& Hartwig, T. 2018, Monthly Notices of the Royal Astronomical Society, 473, 5698

\bibitem[{Dekel \& Silk(1986)}]{Dekel1986}
Dekel, A., \& Silk, J. 1986, Astrophysical Journal, Part 1 (ISSN 0004-637X), vol. 303, April 1, 1986, p. 39-55., 303, 39

\bibitem[{Di~Matteo {et~al.}(2005)Di~Matteo, Springel, \& Hernquist}]{DiMatteo2005}
Di~Matteo, T., Springel, V., \& Hernquist, L. 2005, nature, 433, 604

\bibitem[{Dickey {et~al.}(2019)Dickey, Geha, Wetzel, \& El-Badry}]{Dickey2019}
Dickey, C.~M., Geha, M., Wetzel, A., \& El-Badry, K. 2019, The Astrophysical Journal, 884, 180

\bibitem[{Dubois {et~al.}(2015)Dubois, Volonteri, Silk, Devriendt, Slyz, \& Teyssier}]{Dubois2015}
Dubois, Y., Volonteri, M., Silk, J., {et~al.} 2015, Monthly Notices of the Royal Astronomical Society, 452, 1502

\bibitem[{Ferr{\'e}-Mateu {et~al.}(2021)Ferr{\'e}-Mateu, Durr{\'e}, Forbes, Romanowsky, Alabi, Brodie, \& McDermid}]{Ferre2021}
Ferr{\'e}-Mateu, A., Durr{\'e}, M., Forbes, D.~A., {et~al.} 2021, Monthly Notices of the Royal Astronomical Society, 503, 5455

\bibitem[{Gebhardt {et~al.}(2000)Gebhardt, Bender, Bower, Dressler, Faber, Filippenko, Green, Grillmair, Ho, Kormendy, {et~al.}}]{Gebhardt2000}
Gebhardt, K., Bender, R., Bower, G., {et~al.} 2000, The Astrophysical Journal, 539, L13

\bibitem[{Governato {et~al.}(2010)Governato, Brook, Mayer, Brooks, Rhee, Wadsley, Jonsson, Willman, Stinson, Quinn, {et~al.}}]{Governato2010}
Governato, F., Brook, C., Mayer, L., {et~al.} 2010, nature, 463, 203

\bibitem[{Grcevich \& Putman(2009)}]{Grcevich2009}
Grcevich, J., \& Putman, M.~E. 2009, The Astrophysical Journal, 696, 385

\bibitem[{Greene(2012)}]{Greene2012}
Greene, J.~E. 2012, Nature Communications, 3, 1304

\bibitem[{Greene {et~al.}(2020)Greene, Strader, \& Ho}]{Greene2020}
Greene, J.~E., Strader, J., \& Ho, L.~C. 2020, Annual Review of Astronomy and Astrophysics, 58, 257

\bibitem[{Gutcke {et~al.}(2021)Gutcke, Pakmor, Naab, \& Springel}]{Gutcke2021}
Gutcke, T.~A., Pakmor, R., Naab, T., \& Springel, V. 2021, Monthly Notices of the Royal Astronomical Society, 501, 5597

\bibitem[{Haardt \& Madau(2001)}]{Haardt2001}
Haardt, F., \& Madau, P. 2001, arXiv preprint astro-ph/0106018

\bibitem[{Habouzit {et~al.}(2017)Habouzit, Volonteri, \& Dubois}]{Habouzit2017}
Habouzit, M., Volonteri, M., \& Dubois, Y. 2017, Monthly Notices of the Royal Astronomical Society, 468, 3935

\bibitem[{Haidar {et~al.}(2022)Haidar, Habouzit, Volonteri, Mezcua, Greene, Neumayer, Angl{\'e}s-Alc{\'a}zar, Martin-Navarro, Hoyer, Dubois, {et~al.}}]{Haidar2022}
Haidar, H., Habouzit, M., Volonteri, M., {et~al.} 2022, Monthly Notices of the Royal Astronomical Society, 514, 4912

\bibitem[{Harrington \& Wilson(1950)}]{Harrington1950}
Harrington, R., \& Wilson, A. 1950, Publications of the Astronomical Society of the Pacific, 62, 118

\bibitem[{Hazenfratz {et~al.}(2024)Hazenfratz, Barai, Lanfranchi, \& Caproni}]{Hazenfratz2024}
Hazenfratz, R., Barai, P., Lanfranchi, G.~A., \& Caproni, A. 2024, The Astrophysical Journal

\bibitem[{Hopkins {et~al.}(2014)Hopkins, Kere{\v{s}}, O{\~n}orbe, Faucher-Gigu{\`e}re, Quataert, Murray, \& Bullock}]{Hopkins2014}
Hopkins, P.~F., Kere{\v{s}}, D., O{\~n}orbe, J., {et~al.} 2014, Monthly Notices of the Royal Astronomical Society, 445, 581

\bibitem[{Hoyle \& Lyttleton(1939)}]{hoyle1939}
Hoyle, F., \& Lyttleton, R.~A. 1939, Mathematical Proceedings of the Cambridge Philosophical Society, 35, 405

\bibitem[{Iannuzzi \& Dolag(2011)}]{Iannuzzi2011}
Iannuzzi, F., \& Dolag, K. 2011, Monthly Notices of the Royal Astronomical Society, 417, 2846

\bibitem[{Inayoshi {et~al.}(2016)Inayoshi, Haiman, \& Ostriker}]{Inayoshi2016}
Inayoshi, K., Haiman, Z., \& Ostriker, J.~P. 2016, Monthly Notices of the Royal Astronomical Society, 459, 3738

\bibitem[{Inayoshi {et~al.}(2020)Inayoshi, Visbal, \& Haiman}]{Inayoshi2020}
Inayoshi, K., Visbal, E., \& Haiman, Z. 2020, Annual Review of Astronomy and Astrophysics, 58, 27

\bibitem[{Jardel \& Gebhardt(2012)}]{Jardel2012}
Jardel, J.~R., \& Gebhardt, K. 2012, The Astrophysical Journal, 746, 89

\bibitem[{Kennicutt~Jr(1998{\natexlab{a}})}]{Kennicutt1998star}
Kennicutt~Jr, R.~C. 1998{\natexlab{a}}, Annual Review of Astronomy and Astrophysics, 36, 189

\bibitem[{Kennicutt~Jr(1998{\natexlab{b}})}]{Kennicutt1998global}
---. 1998{\natexlab{b}}, The astrophysical journal, 498, 541

\bibitem[{Kimbrell {et~al.}(2021)Kimbrell, Reines, Schutte, Greene, \& Geha}]{Kimbrell2021}
Kimbrell, S.~J., Reines, A.~E., Schutte, Z., Greene, J.~E., \& Geha, M. 2021, The Astrophysical Journal, 911, 134

\bibitem[{Koudmani {et~al.}(2019)Koudmani, Sijacki, Bourne, \& Smith}]{Koudmani2019}
Koudmani, S., Sijacki, D., Bourne, M.~A., \& Smith, M.~C. 2019, Monthly Notices of the Royal Astronomical Society, 484, 2047

\bibitem[{Koudmani {et~al.}(2022)Koudmani, Sijacki, \& Smith}]{Koudmani2022}
Koudmani, S., Sijacki, D., \& Smith, M.~C. 2022, Monthly Notices of the Royal Astronomical Society, 516, 2112

\bibitem[{Kurapati {et~al.}(2018)Kurapati, Chengalur, Pustilnik, \& Kamphuis}]{Kurapati2018}
Kurapati, S., Chengalur, J.~N., Pustilnik, S., \& Kamphuis, P. 2018, Monthly Notices of the Royal Astronomical Society, 479, 228

\bibitem[{Lanfranchi {et~al.}(2021)Lanfranchi, Hazenfratz, Caproni, \& Silk}]{Lanfranchi2021}
Lanfranchi, G.~A., Hazenfratz, R., Caproni, A., \& Silk, J. 2021, The Astrophysical Journal, 914, 32

\bibitem[{Larson(1974)}]{Larson1974}
Larson, R.~B. 1974, Monthly Notices of the Royal Astronomical Society, 169, 229

\bibitem[{Li {et~al.}(2021)Li, Hammer, Babusiaux, Pawlowski, Yang, Arenou, Du, \& Wang}]{Li2021}
Li, H., Hammer, F., Babusiaux, C., {et~al.} 2021, arXiv preprint arXiv:2104.03974

\bibitem[{Liu {et~al.}(2024)Liu, Veilleux, Canalizo, Tripp, Rupke, Aravindan, Bohn, Hamann, \& Manzano-King}]{Liu2024}
Liu, W., Veilleux, S., Canalizo, G., {et~al.} 2024, The Astrophysical Journal, 965, 152

\bibitem[{Lora {et~al.}(2009)Lora, S{\'a}nchez-Salcedo, Raga, \& Esquivel}]{Lora2009}
Lora, V., S{\'a}nchez-Salcedo, F., Raga, A., \& Esquivel, A. 2009, The Astrophysical Journal, 699, L113

\bibitem[{Maccarone {et~al.}(2005)Maccarone, Fender, \& Tzioumis}]{Maccarone2005}
Maccarone, T.~J., Fender, R.~P., \& Tzioumis, A.~K. 2005, Monthly Notices of the Royal Astronomical Society: Letters, 356, L17

\bibitem[{Magorrian {et~al.}(1998)Magorrian, Tremaine, Richstone, Bender, Bower, Dressler, Faber, Gebhardt, Green, Grillmair, {et~al.}}]{Magorrian1998}
Magorrian, J., Tremaine, S., Richstone, D., {et~al.} 1998, The Astronomical Journal, 115, 2285

\bibitem[{Manni {et~al.}(2015)Manni, Nucita, De~Paolis, Testa, \& Ingrosso}]{Manni2015}
Manni, L., Nucita, A., De~Paolis, F., Testa, V., \& Ingrosso, G. 2015, Monthly Notices of the Royal Astronomical Society, 451, 2735

\bibitem[{Manzano-King {et~al.}(2019)Manzano-King, Canalizo, \& Sales}]{Manzano2019}
Manzano-King, C.~M., Canalizo, G., \& Sales, L.~V. 2019, The Astrophysical Journal, 884, 54

\bibitem[{Mateo(1998)}]{Mateo1998}
Mateo, M. 1998, Annual Review of Astronomy and Astrophysics, 36, 435, \dodoi{10.1146/annurev.astro.36.1.435}

\bibitem[{Mateo {et~al.}(2008)Mateo, Olszewski, \& Walker}]{Mateo2008}
Mateo, M., Olszewski, E.~W., \& Walker, M.~G. 2008, The Astrophysical Journal, 675, 201

\bibitem[{McConnachie(2012)}]{Mcconnachie2012}
McConnachie, A.~W. 2012, The Astronomical Journal, 144, 4

\bibitem[{McKinney \& Gammie(2004)}]{Mckinney2004}
McKinney, J.~C., \& Gammie, C.~F. 2004, The astrophysical journal, 611, 977

\bibitem[{Mezcua(2017)}]{Mezcua2017}
Mezcua, M. 2017, International Journal of Modern Physics D, 26, 1730021

\bibitem[{Mezcua \& S{\'a}nchez(2020)}]{Mezcua2020}
Mezcua, M., \& S{\'a}nchez, H.~D. 2020, The Astrophysical Journal Letters, 898, L30

\bibitem[{Mu{\~n}oz {et~al.}(2018)Mu{\~n}oz, C{\^o}t{\'e}, Santana, Geha, Simon, Oyarz{\'u}n, Stetson, \& Djorgovski}]{Munoz2018}
Mu{\~n}oz, R.~R., C{\^o}t{\'e}, P., Santana, F.~A., {et~al.} 2018, The Astrophysical Journal, 860, 66

\bibitem[{Murray {et~al.}(2005)Murray, Quataert, \& Thompson}]{Murray2005}
Murray, N., Quataert, E., \& Thompson, T.~A. 2005, The Astrophysical Journal, 618, 569

\bibitem[{Murray {et~al.}(2009)Murray, Quataert, \& Thompson}]{Murray2009}
---. 2009, The Astrophysical Journal, 709, 191

\bibitem[{Narayan \& Yi(1995)}]{Narayan1995}
Narayan, R., \& Yi, I. 1995, The Astrophysical Journal, 452, 710

\bibitem[{Nucita {et~al.}(2013)Nucita, Manni, De~Paolis, Vetrugno, \& Ingrosso}]{Nucita2013}
Nucita, A., Manni, L., De~Paolis, F., Vetrugno, D., \& Ingrosso, G. 2013, Astronomy \& Astrophysics, 550, A18

\bibitem[{Nyland {et~al.}(2017)Nyland, Davis, Nguyen, Seth, Wrobel, Kamble, Lacy, Alatalo, Karovska, Maksym, {et~al.}}]{Nyland2017}
Nyland, K., Davis, T.~A., Nguyen, D.~D., {et~al.} 2017, The Astrophysical Journal, 845, 50

\bibitem[{Peebles(1972)}]{Peebles1972}
Peebles, P. 1972, Astrophysical Journal, Vol. 178, pp. 371-376 (1972), 178, 371

\bibitem[{Pillepich {et~al.}(2018)Pillepich, Nelson, Hernquist, Springel, Pakmor, Torrey, Weinberger, Genel, Naiman, Marinacci, {et~al.}}]{Pillepich2018}
Pillepich, A., Nelson, D., Hernquist, L., {et~al.} 2018, Monthly Notices of the Royal Astronomical Society, 475, 648

\bibitem[{Reines \& Comastri(2016)}]{Reines2016}
Reines, A.~E., \& Comastri, A. 2016, Publications of the Astronomical Society of Australia, 33, e054

\bibitem[{Reines {et~al.}(2020)Reines, Condon, Darling, \& Greene}]{Reines2020}
Reines, A.~E., Condon, J.~J., Darling, J., \& Greene, J.~E. 2020, The Astrophysical Journal, 888, 36

\bibitem[{Ricarte {et~al.}(2021)Ricarte, Tremmel, Natarajan, \& Quinn}]{Ricarte2021}
Ricarte, A., Tremmel, M., Natarajan, P., \& Quinn, T. 2021, The Astrophysical Journal Letters, 916, L18

\bibitem[{Sargent {et~al.}(2022)Sargent, Johnson, Reines, Secrest, Van Der~Horst, Cigan, Darling, \& Greene}]{Sargent2022}
Sargent, A.~J., Johnson, M.~C., Reines, A.~E., {et~al.} 2022, The Astrophysical Journal, 933, 160

\bibitem[{Schnittman(2006)}]{Schnittman2006}
Schnittman, J.~D. 2006, arXiv preprint astro-ph/0601406

\bibitem[{Schutte \& Reines(2022)}]{Schutte2022}
Schutte, Z., \& Reines, A.~E. 2022, Nature, 601, 329

\bibitem[{Shakura \& Sunyaev(1973)}]{Shakura1973}
Shakura, N.~I., \& Sunyaev, R.~A. 1973, Astronomy and Astrophysics, 24, 337

\bibitem[{Sharma {et~al.}(2020)Sharma, Brooks, Somerville, Tremmel, Bellovary, Wright, \& Quinn}]{Sharma2020}
Sharma, R.~S., Brooks, A.~M., Somerville, R.~S., {et~al.} 2020, The Astrophysical Journal, 897, 103

\bibitem[{Sharma {et~al.}(2022)Sharma, Brooks, Tremmel, Bellovary, Ricarte, \& Quinn}]{Sharma2022}
Sharma, R.~S., Brooks, A.~M., Tremmel, M., {et~al.} 2022, The Astrophysical Journal, 936, 82

\bibitem[{Shi {et~al.}(2024)Shi, Kremer, \& Hopkins}]{Shi2024}
Shi, Y., Kremer, K., \& Hopkins, P.~F. 2024, arXiv preprint arXiv:2405.17338

\bibitem[{Shin {et~al.}(2019)Shin, Woo, Chung, Baek, Cho, Kang, \& Bae}]{Shin2019}
Shin, J., Woo, J.-H., Chung, A., {et~al.} 2019, The Astrophysical Journal, 881, 147

\bibitem[{Silk(2017)}]{Silk2017}
Silk, J. 2017, The Astrophysical Journal Letters, 839, L13

\bibitem[{Silk \& Rees(1998)}]{Silk1998}
Silk, J., \& Rees, M.~J. 1998, arXiv preprint astro-ph/9801013

\bibitem[{Smith(2021)}]{Smith2021}
Smith, M.~C. 2021, Monthly Notices of the Royal Astronomical Society, 502, 5417

\bibitem[{Somerville(2002)}]{Somerville2002}
Somerville, R.~S. 2002, The Astrophysical Journal, 572, L23

\bibitem[{Spencer {et~al.}(2017)Spencer, Mateo, Walker, \& Olszewski}]{Spencer2017}
Spencer, M.~E., Mateo, M., Walker, M.~G., \& Olszewski, E.~W. 2017, The Astrophysical Journal, 836, 202

\bibitem[{Springel {et~al.}(2005)Springel, Di~Matteo, \& Hernquist}]{Springel2005}
Springel, V., Di~Matteo, T., \& Hernquist, L. 2005, Monthly Notices of the Royal Astronomical Society, 361, 776

\bibitem[{Springel \& Hernquist(2003)}]{Springel2003}
Springel, V., \& Hernquist, L. 2003, Monthly Notices of the Royal Astronomical Society, 339, 289

\bibitem[{Strigari {et~al.}(2007)Strigari, Bullock, Kaplinghat, Diemand, Kuhlen, \& Madau}]{Strigari2007}
Strigari, L.~E., Bullock, J.~S., Kaplinghat, M., {et~al.} 2007, The Astrophysical Journal, 669, 676

\bibitem[{Tchekhovskoy {et~al.}(2011)Tchekhovskoy, Narayan, \& McKinney}]{Tchekhovskoy2011}
Tchekhovskoy, A., Narayan, R., \& McKinney, J.~C. 2011, Monthly Notices of the Royal Astronomical Society: Letters, 418, L79

\bibitem[{Tornatore {et~al.}(2007)Tornatore, Ferrara, \& Schneider}]{Tornatore2007}
Tornatore, L., Ferrara, A., \& Schneider, R. 2007, Monthly Notices of the Royal Astronomical Society, 382, 945

\bibitem[{Volonteri(2010)}]{Volonteri2010}
Volonteri, M. 2010, The Astronomy and Astrophysics Review, 18, 279

\bibitem[{Von~Hoerner(1963)}]{VonHoerner1963}
Von~Hoerner, S. 1963, Z. Astrophys. 57, 47-82, 57

\bibitem[{Walker {et~al.}(2007)Walker, Mateo, Olszewski, Gnedin, Wang, Sen, \& Woodroofe}]{Walker2007}
Walker, M.~G., Mateo, M., Olszewski, E.~W., {et~al.} 2007, The Astrophysical Journal, 667, L53

\bibitem[{White \& Frenk(1991)}]{White1991}
White, S.~D., \& Frenk, C.~S. 1991, Astrophysical Journal, Part 1 (ISSN 0004-637X), vol. 379, Sept. 20, 1991, p. 52-79. Research supported by NASA, NSF, and SERC., 379, 52

\bibitem[{Wiersma {et~al.}(2009)Wiersma, Schaye, \& Smith}]{Wiersma2009}
Wiersma, R.~P., Schaye, J., \& Smith, B.~D. 2009, Monthly Notices of the Royal Astronomical Society, 393, 99

\bibitem[{Yang {et~al.}(2023)Yang, Paragi, Frey, Gurvits, Liao, Liu, Cui, Yang, Chen, Varenius, {et~al.}}]{Yang2023}
Yang, J., Paragi, Z., Frey, S., {et~al.} 2023, Monthly Notices of the Royal Astronomical Society, 520, 5964

\end{thebibliography}
\bibliographystyle{aasjournal}



\end{document}